 \documentclass[final,5p,times,twocolumn,authoryear]{elsarticle}

\usepackage{amssymb}
\usepackage{graphicx}
\usepackage{booktabs}
\usepackage{multirow}
\usepackage{caption}
\usepackage{subcaption}
\usepackage{graphicx}
\usepackage{float}
\usepackage{xcolor}

\begin{document}

\begin{frontmatter}

\title{Edge AI and TinyML for Resource-Constrained Agriculture: Hardware Platforms, Model Optimization, and Deployment Strategies
}

\author[first]{Riya Samanta}
\affiliation[first]{organization={Techno India University},
            state={West Bengal},
            country={India}}

\author[second]{Bidyut Saha}
\affiliation[second]{organization={Independent Researcher},
            state={West Bengal},
            country={India}}

\author[third]{Soumya K Ghosh}
\affiliation[third]{organization={Indian Institute of Technology Kharagpur},
            state={West Bengal},
            country={India}}

\begin{abstract}
Deploying artificial intelligence at the edge is increasingly important for systems that must operate under strict limits on memory, computation, energy, connectivity, and hardware cost. Tiny Machine Learning (TinyML) represents the most resource-constrained regime of this shift, enabling inference on microcontroller-class devices with tightly bounded flash and SRAM budgets. This study examines recent deployment trends in Edge AI and TinyML using precision farming as a representative real-world environment in which sensing nodes, gateways, UAVs, and embedded processors must often operate with intermittent connectivity and limited power. The analysis emphasizes the deployment stack rather than predictive accuracy alone, covering hardware classes, software toolchains, model- and system-level optimization, intelligence placement, and hardware-resource reporting. Across the analyzed studies, microcontroller-class platforms such as ESP32, STM32, and ATmega are prominent endpoint targets, while single-board computers and embedded accelerators support more compute-intensive workloads. Among the  studies included in the optimization comparison, almost 50\% report quantization, making reduced-precision inference the most frequently identified strategy within this subset. In contrast, structured pruning, hardware-aware neural architecture search, and coordinated cross-layer optimization remain comparatively limited. Resource profiling is also inconsistent as flash, peak RAM, MAC count, latency, and energy per inference are not reported uniformly, limiting reproducibility and comparisons across deployments. A recurring systems-level pattern is an inference-training asymmetry in which inference moves toward local devices while training and model lifecycle management remain predominantly centralized. These findings motivate standardized deployment profiling and deeper hardware-software co-design for resource-constrained Edge intelligence for precision farming applications.
\end{abstract}

\begin{keyword}
Edge AI \sep TinyML \sep Resource-constrained computing \sep  Hardware-aware optimization \sep Precision farming
\end{keyword}

\end{frontmatter}

\section{Introduction}
\label{introduction}
Artificial intelligence is increasingly moving from centralized cloud infrastructure toward computation at or near the data source. This shift is driven by applications that require low response latency, reduced communication overhead, improved privacy, and continued operation when network access is intermittent or unavailable \citep{Shi2016Edge,singh2023edge}. However, ``edge'' is not a single computational regime. It spans a continuum from relatively capable gateways, single-board computers, embedded GPUs, and AI accelerators to microcontroller-class endpoints with tightly bounded memory, compute, and energy budgets. Consequently, the central deployment question is no longer only whether a model is accurate, but whether the complete inference pipeline can satisfy the target platform's flash, SRAM, latency, energy, communication, and cost constraints.

Tiny Machine Learning (TinyML) represents the extreme resource-constrained end of this continuum. TinyML systems typically execute on microcontrollers using bare-metal firmware or lightweight real-time operating systems, often without virtual memory or continuous connectivity. Such platforms require compact model representations, statically bounded memory allocation, integer-friendly execution, and careful control of intermediate activation footprints \citep{heydari2025tiny,saha2024tinyml,samanta2025low}. These constraints make deployment a hardware-software co-design problem involving not only model architecture, but also runtime support, compiler behavior, sensing frequency, data movement, communication, and power management.

Precision farming provides a particularly demanding real-world environment in which to examine these deployment constraints. Agricultural systems combine geographically distributed sensing, heterogeneous data modalities, autonomous actuation, intermittent rural connectivity, limited access to power, and strong sensitivity to device and maintenance cost. In India, for example, small and marginal operational holdings constitute approximately $86\%$ of all holdings \citep{PIB_LivelihoodFarmers_2023,AgCensus2015_16_Report}, while infrastructure and digital-access limitations continue to constrain always-connected computational pipelines \citep{GSMA2024}. These conditions make farming systems a useful deployment domain for studying how intelligence can be partitioned across sensing endpoints, local gateways, UAVs, and cloud infrastructure.

Recent farming systems increasingly place inference closer to the sensing source. Distributed sensing nodes with local prediction and edge-gateway coordination have been explored for irrigation control \citep{TinyMLIrrigation}, while quantized and lightweight models enable anomaly detection and vision inference on constrained hardware \citep{Hernandez-Hidalgo2026,Samanta2025}. UAV-assisted systems extend the same principle to mobile edge platforms, where onboard or near-edge processing can reduce communication requirements but introduces additional compute and battery constraints \citep{Liu2021,Annadata2025}. These examples illustrate that deployment feasibility depends strongly on hardware selection, model representation, runtime/toolchain maturity, and intelligence placement rather than on algorithmic accuracy alone \citep{EssanoahArthur2024}.

Model efficiency is therefore only one layer of the problem, but system-level behavior also depends on activation memory, supported operators, generated code size, sampling rate, radio activity, scheduling, and measured energy on the physical target. In this vein, quantization, pruning, lightweight architecture design, knowledge distillation, and hardware-aware neural architecture search are popular ways to reduce parameter storage and computation \citep{Kadhum2025,Koli2025,saha2025gencprunex,saha2025tinytnas} but a model that is compact in file size can still be undeployable if its peak tensor arena exceeds available SRAM or if communication dominates the energy budget.

Another systems-level issue is the emerging asymmetry between inference and training. Inference is progressively decentralized toward endpoint and near-edge devices, while model training, retraining, version control, and lifecycle management remain largely centralized. This architecture improves local autonomy but creates dependencies for model updates, adaptation to environmental drift, and long-term maintenance. Understanding this inference--training decoupling is therefore important for the next generation of communication-efficient, adaptive, and distributed Edge AI systems.

Against this background, the present study analyzes recent Edge AI and TinyML advancements through a system and deployment lens, using precision farming as the application context. The analysis is guided by four research questions:
\begin{enumerate}
    \item \textbf{(RQ1)} Which hardware and software platforms are used across resource-constrained Edge AI and TinyML deployments?
    \item \textbf{(RQ2)} Which model-, compiler-, runtime-, and system-level optimization strategies are used to satisfy memory, latency, and energy constraints?
    \item \textbf{(RQ3)} How is intelligence distributed across endpoint, edge, and cloud tiers, particularly with respect to inference, coordination, and training?
    \item \textbf{(RQ4)} How consistently are deployment-critical resources such as flash, peak RAM, MAC count, latency, power, and energy per inference reported?
\end{enumerate}
The objective of this study is to identify deployment patterns, profiling gaps, and cross-layer design directions that are transferable to resource-constrained Edge intelligence beyond a single agricultural task.

\begin{figure}[!t]
    \centering
    \includegraphics[width=\linewidth]{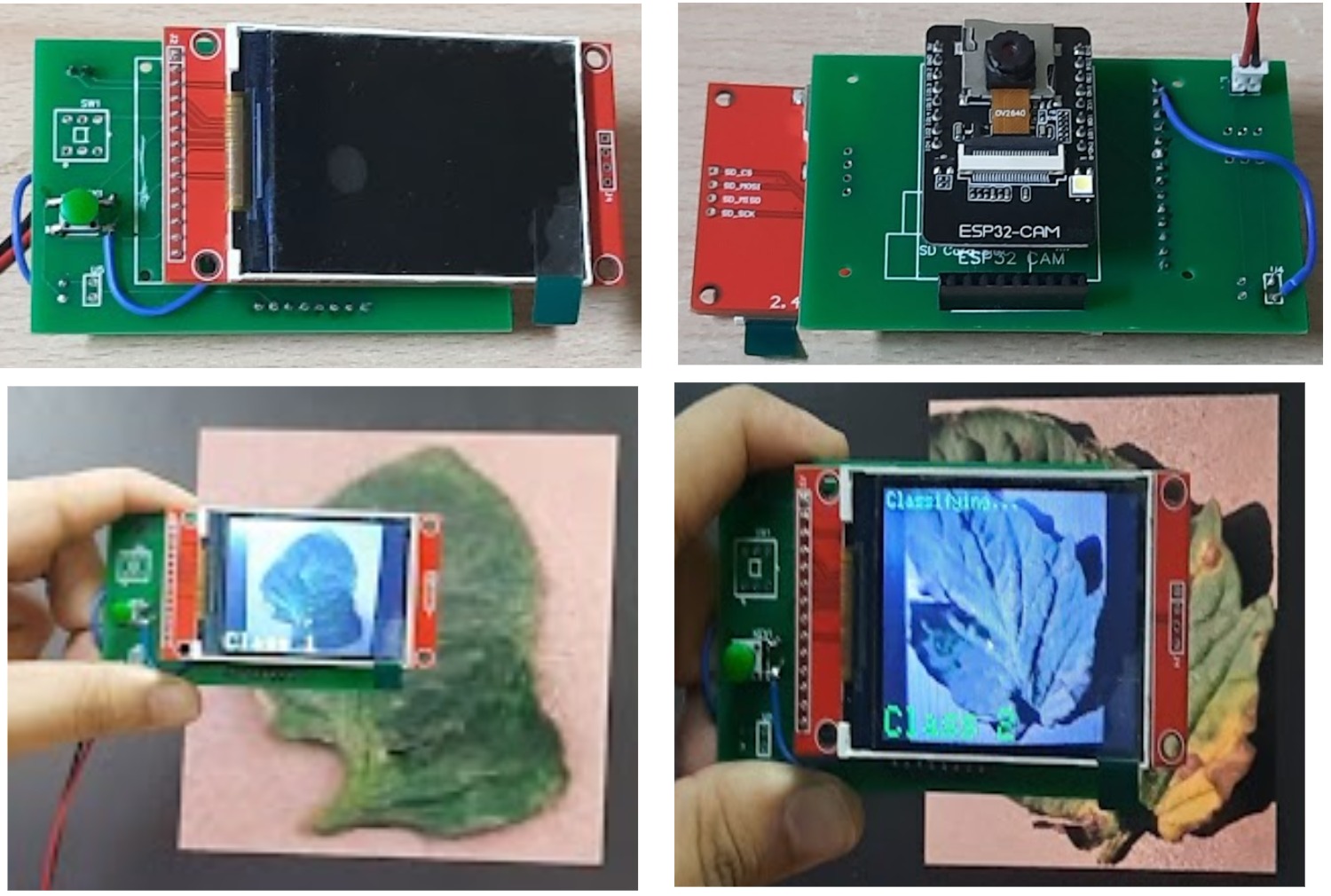}
    \caption{Visual representation of the proposed system uti-
lizing ESP32-CAM, TFT display, and real-time interpreting
tomato plant disease classes (Source: \citep{Samanta2025}).}
    \label{leaf}
    \vspace{-0.2in}
\end{figure}

\section{Edge AI and TinyML as Resource-Constrained Computing Paradigms}
\label{sec:edgeai_tinyml}
\vspace{-0.1in}

This section establishes the deployment stack used to answer \textbf{RQ1} and provides the hardware, runtime, and optimization context needed for \textbf{RQ2}. In particular, it distinguishes the computational regimes encompassed by Edge AI and TinyML before examining the target platforms and toolchains on which the later evidence is interpreted.

Edge Artificial Intelligence (Edge AI) and Tiny Machine Learning
(TinyML) represent a shift from centralized, cloud-based machine
learning toward decentralized, on-device intelligence. By performing
inference locally rather than transmitting raw data to remote servers,
these paradigms reduce dependence on persistent connectivity and
centralized infrastructure \citep{singh2023edge,heydari2025tiny},
which is particularly relevant in distributed and resource-constrained
agricultural environments. From a systems perspective, deployment can be viewed as a stack spanning the sensing interface, target processor, runtime/compiler, model representation, communication layer, and lifecycle-management tier. Bottlenecks at any one of these layers can determine whether a nominally lightweight model is actually deployable on the physical target.

\vspace{-0.05in}
\subsection{Edge AI}
\vspace{-0.05in}
Edge AI refers to the deployment of machine learning models on
computing platforms located at or near the edge of the network,
including embedded processors, edge gateways, single-board computers,
smartphones, UAV computers, and dedicated AI accelerators
\citep{singh2023edge}. In time-critical and geographically distributed
applications, Edge AI reduces communication overhead and latency,
improves data privacy, and enhances system responsiveness by placing
computation closer to the data source \citep{abou2025edge}.

In precision farming, Edge AI enables localized decision-making for
irrigation control, anomaly detection, crop health monitoring, pest
identification, UAV analytics, and multi-sensor fusion without
requiring continuous cloud interaction. This improves operational
resilience under intermittent connectivity and reduces the bandwidth
required for transmitting high-volume sensor, image, or video data.
However, Edge AI platforms vary substantially in processing capacity,
memory availability, power demand, operating-system support, and
deployment cost. Consequently, the term encompasses both moderately
constrained embedded systems and comparatively capable platforms such
as single-board computers and embedded GPUs.

\subsection{TinyML}

TinyML, or Tiny Machine Learning, is a specialized subset of Edge AI
that focuses on ultra-low-power and memory-constrained devices,
typically microcontrollers \citep{heydari2025tiny}. In contrast to
conventional machine-learning systems built on desktop CPUs, GPUs, or
cloud servers, TinyML models are optimized for small memory footprints,
low-latency inference, and strict power consumption under highly
constrained hardware budgets
\citep{saha2024tinyml,samanta2025low}.

TinyML systems commonly operate using bare-metal firmware or lightweight
real-time operating systems rather than general-purpose operating
systems. They usually execute within statically bounded flash and SRAM
budgets and may lack virtual memory, high-capacity storage, or continuous
network access. These constraints motivate the use of model compression,
integer quantization, lightweight architecture design, pruning,
knowledge distillation, and hardware-aware optimization
\citep{saha2025gencprunex,saha2025efficiency,saha2025tinytnas}.

Although all TinyML systems fall within the broader Edge AI ecosystem,
Edge AI may involve more capable processors, gateways, single-board
computers, embedded GPUs, or dedicated accelerators. TinyML explicitly
emphasizes deployment on microcontroller-class hardware with stringent
memory, compute, and energy limitations \citep{abou2025edge}. The
distinction is therefore not determined only by where inference is
performed, but also by the available memory hierarchy, processing
capability, execution environment, power envelope, operator support,
and deployment toolchain.

Although this conceptual distinction is important, deployment
feasibility is ultimately determined by the interaction among the
target hardware, software framework, model architecture, and
optimization strategy. In precision farming, these choices must also
account for intermittent connectivity, environmental exposure,
restricted power supply, maintenance limitations, and the cost
sensitivity of smallholder-oriented deployments.

\subsection{Hardware Platforms for Resource-Constrained Edge AI and TinyML}
\label{subsec:hardware_platforms}

The hardware ecosystem for resource-constrained Edge AI and TinyML spans a
broad range of computational capability, from ultra-low-power
microcontrollers to single-board computers, embedded GPUs, and
dedicated neural accelerators. Selection of an appropriate platform
depends on the sensing modality, model complexity, latency requirement,
energy availability, communication infrastructure, deployment cost,
and expected maintenance burden.

\subsubsection{Microcontroller-Class Platforms}

Microcontrollers constitute the principal deployment targets for
TinyML because they offer low cost, compact form factor, low standby
power, and direct interfacing with agricultural sensors and actuators.
Representative platforms reported in the included studies include
ESP32, ESP32-CAM, ESP-EYE, STM32, ATmega, and Arduino Nano 33
BLE-class boards.

ESP32-class platforms are particularly attractive for low-cost
precision farming because they combine wireless connectivity,
general-purpose input/output interfaces, analogue-to-digital
conversion, and sufficient processing capability for compact
machine-learning inference. Camera-enabled variants such as ESP32-CAM
and ESP-EYE support image-based tasks including plant disease
classification and crop monitoring. However, their limited internal
SRAM constrains input resolution, intermediate feature-map size, and
model depth. Deployment therefore commonly requires reduced-resolution
inputs, integer quantization, lightweight convolutional architectures,
and careful management of activation buffers.

Arduino Nano 33 BLE-class boards and related Arm Cortex-M platforms
are well suited to low-dimensional sensing applications such as soil
condition classification, environmental anomaly detection, acoustic
pest monitoring, and irrigation control. Their integrated sensors and
low-power wireless interfaces simplify rapid prototyping. However,
frequent wireless transmission may consume more energy than local
inference, particularly when the node sends raw or high-frequency
sensor data.

STM32 microcontrollers offer a broader range of processor, memory, and
power configurations. Cortex-M4, Cortex-M7, and Cortex-M33 devices
provide digital signal-processing instructions and, in selected
variants, hardware features that accelerate multiply--accumulate
operations. These platforms are suitable for more demanding
agricultural workloads, including compact vision models, low-power
weed detection, and multi-sensor fusion. Their deployment can benefit
from vendor-specific optimization libraries and code-generation tools,
although they generally require greater embedded-system expertise than
high-level prototyping platforms.

ATmega-class microcontrollers remain useful for highly compact
classical machine-learning models, including decision trees, compressed
random forests, and rule-based systems. Their primary advantages are
very low cost, simplicity, and broad availability. However, their
limited SRAM, low processor frequency, and lack of dedicated
machine-learning acceleration restrict their suitability for deep
neural networks and image-intensive applications.

\subsubsection{Single-Board Computers and Edge Gateways}

Single-board computers such as Raspberry Pi-class platforms provide
substantially greater memory and processing capacity than
microcontrollers. They support Linux-based operating environments,
local databases, high-level machine-learning runtimes, and simultaneous
communication with multiple field nodes.

In precision farming, these platforms can function as farm-level
gateways that aggregate sensor data, perform multi-sensor fusion,
coordinate irrigation decisions, maintain local records, and execute
crop-health analytics. They are also suitable for applications
requiring larger neural networks, multiple cameras, graphical user
interfaces, or local storage.

Their flexibility is accompanied by higher energy demand, more complex
software maintenance, and dependence on operating-system and storage
reliability. They typically consume several watts and may require
larger batteries or solar-power systems. Consequently, they are more
appropriate as shared gateways or near-edge processing nodes than as
densely distributed battery-powered field sensors.

\subsubsection{Embedded GPUs and AI Accelerators}

Embedded GPU platforms, including Jetson-class devices, are appropriate
for computationally intensive agricultural tasks such as real-time
object detection, UAV image analysis, robotic navigation, weed
localization, and multimodal sensor fusion. These devices support
larger models and accelerator-oriented runtimes, enabling higher
throughput than microcontroller-class hardware.

However, their increased processing capability is associated with
higher cost, thermal requirements, and multi-watt power consumption.
These characteristics reduce their suitability for unattended,
low-cost sensing nodes but make them useful for UAVs, agricultural
robots, greenhouse controllers, and local edge stations.

Dedicated neural accelerators, including Edge TPU-class devices, can
improve operations-per-watt for supported quantized models. They are
useful when compute-intensive inference must be offloaded from a
microcontroller or single-board computer. Nevertheless, accelerator
deployment introduces restrictions related to supported operators,
required numerical precision, model-conversion compatibility, and
hardware availability. For smallholder-oriented systems, their
benefits should therefore be evaluated at the complete system level
rather than only through inference throughput.

\subsection{Software Frameworks and Deployment Toolchains}
\label{subsec:tinyml_toolchains}

A deployment-oriented TinyML workflow involves more than training a
compact model. It typically includes data acquisition, preprocessing,
model compression, graph conversion, hardware-specific compilation,
firmware integration, memory allocation, device programming, and
on-target profiling. The maturity of the software toolchain strongly
influences whether a model can progress from laboratory evaluation to
reliable on-field deployment.

TensorFlow Lite for Microcontrollers is one of the most widely used
frameworks for executing neural networks on microcontroller-class
hardware \citep{warden2019tinyml,david2021tensorflow}. Models are generally trained using TensorFlow or Keras,
converted to a lightweight representation, quantized, embedded into
firmware, and executed through a statically allocated tensor arena.
Static allocation is important because many microcontrollers do not
provide virtual memory and offer limited support for dynamic memory
management.

Its compact runtime is suitable for offline sensing and classification,
although the model must use operators supported by the target runtime.
The tensor arena, model parameters, input buffers, and application
firmware must also fit within the available SRAM and flash memory.

TensorFlow Lite is more suitable for Linux-capable Edge AI platforms,
including Raspberry Pi-class gateways, smartphones, and embedded
computers. It supports a broader operator set and more flexible
execution than its microcontroller-oriented counterpart, enabling
larger crop-classification, pest-detection, and sensor-fusion models.

Edge Impulse \citep{hymel2022edge} provides an integrated workflow covering data acquisition,
feature extraction, model training, quantization, profiling, and
embedded deployment. Its support for sensor, audio, and vision
applications makes it useful for rapid agricultural prototyping. It
can estimate flash usage, RAM demand, and inference latency before
deployment and export firmware libraries for multiple hardware
families.

However, tool-generated estimates should still be verified on the
physical target device. Memory allocation, peripheral activity,
communication interfaces, camera capture, sensor polling, and
environmental conditions may alter actual deployment behaviour.

STM32Cube.AI \citep{stm32cubeai} supports conversion and optimization of trained neural
networks for STM32 microcontrollers. In combination with optimized
libraries, it can exploit Arm Cortex-M instructions
for convolution, activation, pooling, and fully connected operations.
Such hardware-specific optimization can reduce latency and memory
overhead for continuous agricultural sensing applications.

For embedded GPU platforms, TensorRT \citep{nvidia_tensorrt} performs graph optimization,
operator fusion, kernel selection, and reduced-precision execution.
It is particularly relevant to Jetson-based plant disease detection,
UAV analytics, and agricultural robotics. ONNX \citep{chen2018tvm} can serve as an
intermediate representation when transferring models between training
frameworks and deployment runtimes, although operator compatibility
and quantization behaviour must be validated carefully.

Compiler-oriented frameworks such as Apache TVM \citep{chen2018tvm} and microTVM \citep{liu2023deploying} provide
hardware-specific code generation, operator scheduling, and
cross-compilation for embedded targets. These tools can support
fine-grained profiling and hardware-aware optimization but generally
require greater expertise in compiler configuration, firmware
integration, and embedded debugging.

For deployment-oriented comparison, a software framework should be
evaluated not only by whether it converts a model successfully, but
also by its supported operators, quantization modes, generated code
size, peak tensor memory, availability of optimized kernels, processor
compatibility, profiling support, firmware-integration complexity,
licensing requirements, and model-update mechanisms.

The toolchain therefore acts as an optimization layer in its own right. Graph rewriting, kernel selection, operator fusion, static memory planning, target-specific code generation, and integer-kernel availability can change measured latency and memory usage even when the high-level neural architecture is unchanged. For this reason, deployment claims should identify not only the model and hardware but also the conversion path, runtime, compiler or vendor library, numerical precision, and whether resource measurements were obtained from estimates or from the physical target.

\subsection{Optimization Foundations for Resource-Constrained Edge AI and TinyML}
\label{subsec:model_optimization}

Models developed on desktop CPUs or GPUs are rarely suitable for direct
deployment on microcontrollers. Their parameter storage, activation
memory, computational complexity, and energy demand may exceed the
available flash, SRAM, latency, or power budget. Model compression and
hardware-aware optimization are therefore central to resource-constrained
TinyML deployment. 


\subsubsection{Quantization}

Quantization \citep{lamaakal2025tiny} reduces the numerical precision of model weights and
activations, commonly from 32-bit floating-point values to 8-bit
integers. INT8 quantization can reduce parameter storage by
approximately four times and may also reduce memory-access overhead,
inference latency, and energy consumption when integer-optimized
kernels are available.

Post-training quantization is attractive because it can be applied
after model training with limited modification to the original
pipeline. However, its effectiveness depends on the
representativeness of the calibration dataset. In precision farming,
calibration samples should capture realistic variation in
illumination, crop growth stage, sensor range, soil condition,
background complexity, and environmental noise. Quantization
calibrated using narrowly controlled data may preserve benchmark
accuracy while degrading performance in field conditions.

Quantization-aware training simulates reduced-precision arithmetic
during model training and generally provides better accuracy
retention for sensitive models. Mixed-precision approaches may
assign different bit widths to different layers, although their
practical benefit depends on support from the target processor,
runtime, and compiler. Extremely low-precision representations, such
as 4-bit or binary models, can further reduce memory use but may
require specialized hardware kernels and can result in substantial
accuracy loss.

\subsubsection{Pruning}

Pruning \citep{saha2025gencprunex} removes redundant weights, filters, channels, or layers from
a trained model. Unstructured pruning removes individual weights and
creates sparse matrices. However, its practical benefit may remain
limited on microcontrollers that lack efficient sparse-execution
support. A reduction in the number of non-zero parameters does not
necessarily produce a proportional reduction in latency or energy.

Structured pruning removes complete filters, channels, or
computational blocks and therefore produces smaller dense networks
that can be executed using conventional embedded kernels. For
agricultural vision models, channel-wise pruning can reduce both
parameter storage and intermediate activation memory. This is
particularly important because peak activation memory, rather than
model-file size alone, frequently determines whether a convolutional
network can execute on a microcontroller. The effectiveness of pruning should therefore be evaluated through
on-device flash usage, peak SRAM, latency, MAC operations, and
energy per inference rather than only through the percentage of
removed parameters or the achieved sparsity ratio.

\subsubsection{Knowledge Distillation}

Knowledge distillation \citep{moslemi2024survey} trains a compact student model to reproduce
the behaviour of a larger teacher model. Instead of learning only
from class labels, the student also learns from the probability
distribution or intermediate features generated by the teacher. This
can preserve part of the predictive capability of the larger model
without increasing the inference-time complexity of the student.

In precision farming, a high-capacity teacher model may be trained
using high-resolution crop imagery, multispectral data, or multimodal
environmental measurements, while the smaller student model is
deployed on an MCU or farm gateway. Distillation may be particularly
useful for distinguishing visually similar disease symptoms or
overlapping irrigation and soil-condition patterns.

However, distillation does not automatically guarantee deployment
feasibility. The student architecture must still satisfy the flash,
SRAM, latency, power, and operator constraints of the target
platform. Knowledge distillation is therefore most effective when
combined with quantization, pruning, or lightweight architecture
design.

\subsubsection{Lightweight Model Architectures}

Lightweight neural networks reduce computational demand using
efficient building blocks such as depthwise separable convolution,
inverted residual structures, grouped convolution, compact
bottlenecks, and channel-shuffling operations. MobileNet \citep{sandler2018mobilenetv2},
EfficientNet-Lite, ShuffleNet, SqueezeNet \citep{iandola2016squeezenet}, and custom shallow CNNs
illustrate this design strategy.

However, a model described as lightweight for a smartphone or
embedded GPU may remain too large for a microcontroller. Agricultural
TinyML systems frequently require task-specific models with reduced
input resolution, fewer channels, shallow feature extractors, and
compact classification heads.

For low-dimensional soil, weather, and irrigation data, classical
machine-learning methods such as decision trees, random forests,
gradient boosting, or linear models may offer a more suitable
accuracy--resource trade-off than deep neural networks. Their smaller
memory requirements, simpler execution, and greater interpretability
may also facilitate maintenance and validation in field deployments.

\subsubsection{Hardware-Aware Optimization}

Hardware-aware optimization incorporates target-device constraints
during model design rather than applying compression only after
training. The optimization objective may jointly consider predictive
accuracy, model size, flash usage, peak SRAM, MAC count, measured
latency, and energy per inference. Hardware-aware neural architecture search \citep{saha2025tinytnas,fayyazi2026marco} can identify model
configurations that satisfy platform-specific constraints by varying
layer depth, channel width, kernel size, input resolution, operator
type, and quantization precision. Device-in-the-loop evaluation is
particularly valuable because parameter count and MAC count do not
always correspond directly to measured latency or energy on real
hardware.

For precision agriculture, hardware-aware optimization should also
consider the sensing and communication pipeline. Reducing image
resolution, sampling frequency, inference frequency, or radio
transmission rate may yield greater system-level energy savings than
model compression alone. Efficient TinyML deployment should therefore
be treated as a cross-layer co-design problem involving sensing,
preprocessing, model execution, communication, scheduling, and power
management.


\subsubsection{System-Level and Cross-Layer Optimization}
At system level, the dominant energy or latency cost may lie outside the neural network itself. Sensor sampling, camera capture, preprocessing, memory movement, radio transmission, wake/sleep scheduling, and actuator control can all determine the end-to-end resource budget. Consequently, resource-constrained Edge AI should optimize the complete sensing--compute--communication path. Adaptive sampling, event-triggered inference, communication-aware offloading, dynamic voltage and frequency scaling (DVFS), and workload scheduling can complement model compression, and in some deployments may yield larger system-level savings than reducing parameter count alone.

\begin{figure*}[!t]
    \centering
    \includegraphics[width=0.8\linewidth]{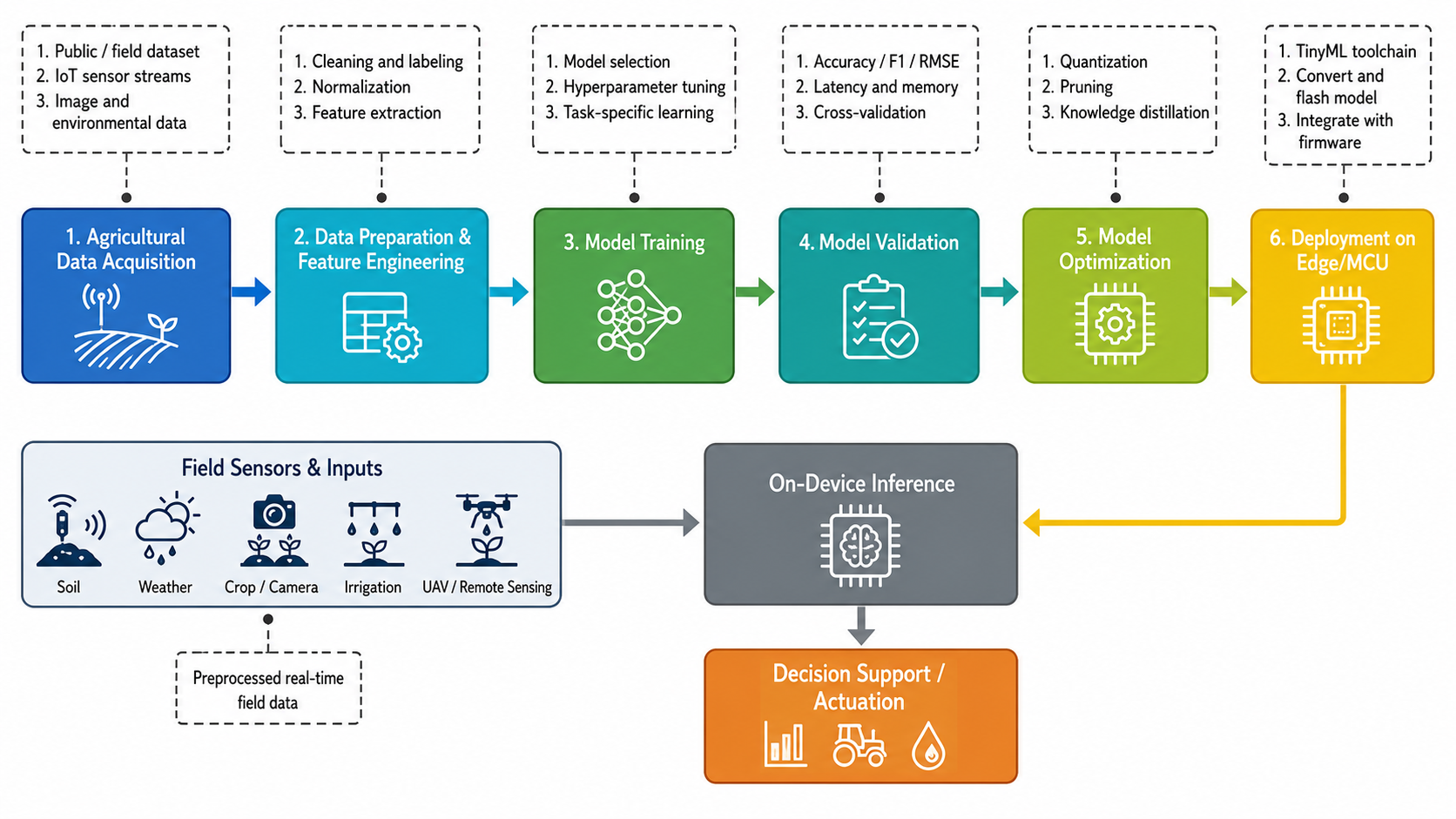}
\caption{Functional taxonomy and end-to-end design workflow of an Edge AI or TinyML system for precision farming.}
    \label{fig:tinyml_precision_farming_workflow}
\end{figure*}

\subsection{End-to-End Deployment Workflow}

A deployment-oriented Edge AI or TinyML system for precision farming
can be organized into the following stages:

\begin{enumerate}
    \item \textbf{Data acquisition and preprocessing:}
    Agricultural data are collected from cameras, environmental
    sensors, soil probes, acoustic devices, or UAV systems. The data
    should represent the intended crop, region, season, device, and
    environmental conditions.

    \item \textbf{Model development and evaluation:}
    An appropriate machine-learning architecture is trained and
    evaluated using both predictive and deployment-related criteria.

    \item \textbf{Model optimization:}
    Quantization, pruning, knowledge distillation, lightweight
    architecture design, or hardware-aware search is applied to
    satisfy target resource constraints.

    \item \textbf{Model conversion and compilation:}
    The optimized model is converted into a hardware-compatible
    representation and compiled using the selected runtime,
    vendor-specific library, or accelerator toolchain.

    \item \textbf{On-device inference and profiling:}
    The model is executed on the physical device, and flash usage,
    peak RAM, inference latency, energy per inference, average device
    power, and communication overhead are measured.

    \item \textbf{Application integration and lifecycle management:}
    Inference outputs are connected to alerts, displays, pumps,
    relays, irrigation controllers, UAV navigation, or farm
    management systems. Model updates, sensor recalibration, fault
    handling, environmental protection, and long-term maintenance
    must also be considered.
\end{enumerate}

The workflow in Figure~\ref{fig:tinyml_precision_farming_workflow}, demonstrates that TinyML deployment is a
hardware--software co-design problem rather than merely the execution
of a small machine-learning model. In precision farming, the final
system must remain reliable under changing weather, sensor drift,
dust, moisture, limited connectivity, constrained power supply, and
restricted maintenance access. Accordingly, the selection of
hardware, runtime, compression strategy, and communication mode
should be based on the complete agricultural use case rather than
predictive accuracy alone.  Additionally, a
concise technical distinction between TinyML and broader Edge AI in
precision farming is presented in Table~\ref{tab:tinyml_edgeai}.

\begin{table*}[!t]
\centering
\caption{Distinction between TinyML and broader Edge AI.}
\vspace{-0.05in}
\label{tab:tinyml_edgeai}
\small
\begin{tabular}{p{2.8cm}p{6.1cm}p{6.1cm}}
\hline
\textbf{Dimension} & \textbf{TinyML} & \textbf{Edge AI} \\
\hline
Typical hardware &
Microcontrollers and low-power embedded controllers &
SBCs, gateways, smartphones, embedded GPUs, and AI accelerators \\

Memory &
Typically a few KB to a few MB &
Typically hundreds of MB to several GB \\

Execution environment &
Bare-metal firmware, lightweight runtime, or RTOS &
Linux, general-purpose operating systems, or accelerator-specific runtimes \\

Power profile &
Usually mW-scale, battery-powered, or duty-cycled &
Generally higher, often ranging from sub-watt to several watts \\

Connectivity &
Can operate fully offline at the sensing endpoint &
May operate offline or within edge--cloud and device--gateway architectures \\

Precision farming tasks &
Sensor anomaly detection, compact classification, irrigation control, and local decision-making &
Object detection, UAV analytics, multimodal sensor fusion, robotic navigation, and gateway coordination \\

Main constraints &
Flash memory, SRAM, computation, and battery life &
Latency, throughput, energy, cost, thermal load, and communication overhead \\

Deployment platforms &
ESP32, STM32, ATmega, and Arduino Nano 33 BLE &
Jetson Nano, edge gateways, SBCs, and higher-capability UAV edge systems \\
\hline
\end{tabular}
\end{table*}

\vspace{-0.15in}
\section{Study Scope and Evidence Collection}
\label{meth}
\vspace{-0.1in}

This study presents an analysis of research at the overlap of resource-constrained edge intelligence, TinyML, energy-efficient computing, and precision agriculture. Studies were identified through bibliographic databases, scholarly search engines, and publisher platforms, including \textit{Google Scholar, IEEE Xplore, ScienceDirect, SpringerLink, ACM Digital Library}, and \textit{Scopus}, to ensure extensive coverage of peer-reviewed journal and conference articles.

A structured search strategy involved structured Boolean search queries such as “Precision agriculture” AND “TinyML” AND “Energy efficiency,” “Low-cost precision farming” AND (“Edge AI” OR “Machine learning”) AND “Resource-constrained devices,” “Smart farming” AND “Low-power computing” AND “Model optimization techniques,” and “Affordable precision farming” AND “Hardware-aware model optimization” AND (“Energy-aware systems” OR “Sensor fusion”). 

Although edge computing and embedded machine learning predate 2020, the dedicated TinyML ecosystem and MCU-oriented deployment toolchains matured substantially around 2019–2021 \citep{warden2019tinyml}. Accordingly, the primary evidence window emphasizes 2020–2026, while earlier foundational works are retained where necessary to establish the architectural and methodological background.

\begin{table*}[!t]
\centering
\caption{Application-wise deployment evidence for Edge AI and TinyML in precision farming}
\vspace{-0.1in}
\label{tab:application_summary}
\footnotesize
\setlength{\tabcolsep}{3.5pt}
\renewcommand{\arraystretch}{1.3}

\resizebox{\textwidth}{!}{
\begin{tabular}{llllllll}
\toprule
\textbf{Application} &
\textbf{References} &
\textbf{Hardware} &
\textbf{Model} &
\textbf{Optimization} &
\textbf{Validation Status} &
\textbf{Key Metric} &
\textbf{Offline} \\
\midrule

\multirow{8}{*}{\textbf{Plant Disease Detection}}
& \citep{Samanta2025}
& ESP32-CAM
& CNN
& Quantization
& Physical
& 92\% Acc.
& Yes \\

& \citep{Koli2025}
& Edge Device
& DS-CNN
& Lightweight Arch.
& Physical
& 96\% Acc.
& Yes \\

& \citep{Gookyi2024}
& Arduino BLE 33
& CNN
& Augmentation + Tuning
& Physical
& 94.6\%, 7.6 ms
& Yes \\

& \citep{Annadata2025}
& ESP-EYE 32
& CNN
& Path Optimization
& Physical
& 90\% Acc.
& Yes \\

& \citep{tao2025icrop+}
& LoRa + Edge
& CNN
& Adaptive Offloading
& Physical
& --
& Partial \\

& \citep{Kouzinopoulos2025}
& STM32U575
& YOLOv8n
& Pruning + INT8
& Physical
& 51.8 mJ/inf.
& Yes \\

& \citep{Madiwal2025}
& Edge Device
& MobileNet/EfficientNet
& Quantization
& Review
& --
& Partial \\

& \citep{tasci2026dbla}
& NVIDIA Jetson Nano
& DBLA-MobileNetV2
& FP16 + TensorRT
& Physical
& 97.90\% Prec., 12.40 FPS
& Yes \\

\midrule

\multirow{3}{*}{\textbf{Precision Irrigation}}
& \citep{sustainable}
& ESP32
& Gradient Boosting
& Model Conversion
& Physical
& MAPE $<$ 1\%
& Yes \\

& \citep{Hernandez-Hidalgo2026}
& MCU
& Quantized NN
& 8-bit Quantization
& Physical
& 6.2 KB Model
& Yes \\

& \citep{Thirumalaiah2025}
& LoRa + Edge
& CNN + Analytics
& Arch. Design
& Conceptual
& Water Savings
& Yes \\

\midrule

\multirow{1}{*}{\textbf{Soil Monitoring}}
& \citep{Bhattacharya2024}
& MCU + Blockchain
& TinyML
& DVFS + GA Scheduling
& Simulated
& 8.5\% Energy $\downarrow$
& Yes \\

\midrule

\multirow{1}{*}{\textbf{Crop Recommendation}}
& \citep{Baishya2025}
& ATMega328P
& Random Forest
& Model Compression
& Physical
& 99\% Size Reduction
& Yes \\

\midrule

\multirow{3}{*}{\textbf{UAV / Distributed}}
& \citep{Soltani2025}
& UAV + Edge
& Split Learning
& Energy-aware Training
& Simulated
& 86\% Energy $\downarrow$
& No \\

& \citep{Annadata2025}
& ESP-EYE 32
& CNN
& Path Optimization
& Physical
& 90\% Acc.
& Yes \\

& \citep{Hayajneh2024}
& UAV + Edge MCU
& CNN (Transfer Learning)
& Transfer Learning + Quantization
& Simulated
& Improved Inference Efficiency
& Yes \\

\bottomrule
\end{tabular}
}
\vspace{-0.2in}
\end{table*}

\subsection{Study Eligibility and Evidence Classification}

To distinguish demonstrated deployment from preliminary or conceptual
feasibility, the included studies were classified into three evidence
categories:

\begin{enumerate}
    \item \textbf{Physical hardware deployment:} The model or system was
    executed on the stated physical edge device, microcontroller, UAV
    computer, single-board computer, or gateway, with device-level
    measurements such as memory usage, latency, energy consumption, or
    throughput.

    \item \textbf{Experimental or simulation-based evaluation:} The
    proposed approach was evaluated through simulation, emulation,
    controlled testbed experimentation, development-board benchmarking,
    or analytical comparison, but complete operation on the claimed
    agricultural hardware or in a field environment was not demonstrated.

    \item \textbf{Conceptual, architectural, or review-level contribution:}
    The study proposed an architecture, framework, taxonomy, design
    methodology, or review synthesis without reporting a complete
    implementation on physical agricultural hardware.
\end{enumerate}

Studies were included when they contributed deployment-relevant evidence
in at least one of these three forms: physical hardware validation,
experimental or simulation-based feasibility evaluation, or a clearly
specified conceptual architecture relevant to agricultural edge
deployment. The evidence category was recorded separately to prevent
conceptual or simulated systems from being interpreted as validated
hardware deployments.

The additional inclusion criteria were: (i) relevance to precision
agriculture, smart farming, or agricultural monitoring and control;
(ii) use or analysis of Edge AI, TinyML, embedded intelligence, edge
gateways, or UAV-assisted edge computing; (iii) consideration of at
least one deployment-related factor, such as flash memory, RAM, model
size, computational complexity, latency, energy consumption,
connectivity, hardware cost, or deployment architecture; and
(iv) publication as a peer-reviewed journal article, refereed conference
paper, or sufficiently detailed archival technical report.

Studies were excluded when they: (i) addressed only cloud-based
agricultural AI without any edge, embedded, or near-edge component;
(ii) focused exclusively on algorithmic accuracy without discussing
deployment feasibility; (iii) lacked sufficient methodological or
technical information to determine the proposed hardware, architecture,
or evaluation setting; or (iv) were non-archival opinion articles,
editorials, abstracts, or promotional descriptions.

The initial search yielded 45 candidate articles. After screening for relevance, 34 papers underwent full-text assessment, of which 28 satisfied the inclusion criteria.


Of the 28 included studies, 15 provided sufficiently detailed application-level deployment evidence for inclusion in Table \ref{tab:application_summary}, 13 reported identifiable optimization strategies for inclusion in Table \ref{tab:optimization_strategies}, and 10 provided at least one explicit hardware-resource or energy metric for inclusion in Table \ref{tab:energy_resource}. These subsets overlap and should not be interpreted as separate study populations.

\begin{figure}[!t]
    \centering
    \includegraphics[width=\linewidth]{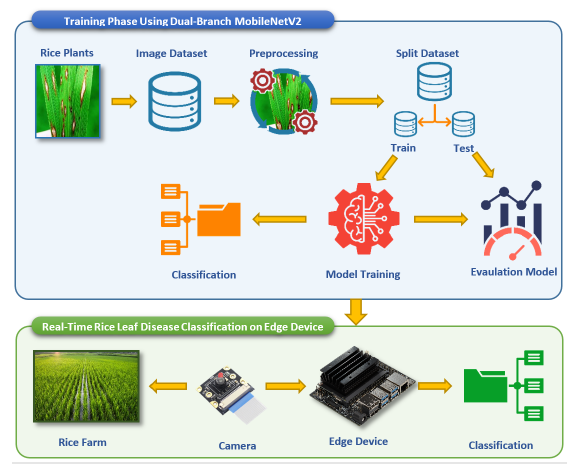}
    \caption{Overview of the methodology, illustrating the
training phase with DBLA-MobileNetV2 and the deployment
phase on Jetson Nano for real-time rice leaf disease classifi-
cation (Source: \citep{tasci2026dbla}).}
    \label{dvla}
    \vspace{-0.2in}
\end{figure}

\vspace{-0.15in}
\section{Application Landscape of Edge AI and TinyML in Precision Agriculture}
\vspace{-0.1in}

Previous application-centric works organize smart agriculture around agronomic stages or use cases. Here, the same application categories are retained, but they are interpreted as distinct Edge AI workload classes with different sensing modalities, computational intensity, memory behavior, communication requirements, and target hardware. This allows the application evidence to be read through a deployment-systems lens without discarding the domain-specific structure of the literature.

These application-level deployments form the common empirical basis for all four research questions: their hardware and software targets inform \textbf{RQ1}, their optimization choices inform \textbf{RQ2}, their execution and coordination patterns inform \textbf{RQ3}, and their reported memory, latency, and energy measurements inform \textbf{RQ4}. Subsequent sections reorganize the same evidence along these systems dimensions rather than treating the application categories as independent end points of the analysis.

The deployment of Edge AI and TinyML in precision farming has accelerated in response to the need for low-cost, low-power, and connectivity-resilient intelligent systems \citep{saha2024tinyml,saha2025efficiency,samanta2025low}. As summarized in Table~\ref{tab:application_summary}, the five retained domains that are \textit{disease detection, precision irrigation, soil monitoring, crop recommendation, and UAV-assisted distributed intelligence}, map to markedly different computational workloads, ranging from image-intensive CNN inference to low-dimensional tabular prediction and distributed mobile-edge processing. The validation status distinguishes physical hardware execution from simulation-based, conceptual, and secondary-study evidence; this distinction is important because benchmark accuracy alone does not establish deployability on the claimed target platform.

\subsection{Plant Disease Detection}

 From an Edge AI systems perspective, plant disease detection is primarily a constrained vision workload. It combines image acquisition, preprocessing, convolution-heavy inference, and often local display or alert generation, making peak activation memory, input resolution, MAC count, and camera-buffer management central deployment constraints. The application is therefore useful for examining the boundary between MCU-class TinyML vision and higher-capability embedded edge inference, particularly under field-induced distribution shift in illumination, background, and viewpoint.
 
 Plant diseases represent the most mature application of TinyML in the agriculture sector. Convolutional neural networks (CNNs) have been deployed on microcontroller-class platforms including ESP32-CAM \citep{Samanta2025}, Arduino Nano BLE 33 \citep{Gookyi2024}, ESP-EYE \citep{Annadata2025}, and STM32U575 \citep{Kouzinopoulos2025}, as well as higher-resource edge devices such as Jetson Nano using optimized MobileNet variants with attention mechanisms \citep{tasci2026dbla}, refer Fig.~\ref{dvla}. Reported classification accuracies range from 90\% to 96\% (refer to Table~\ref{tab:application_summary}), with selected works providing hardware-level metrics such as 7.6\,ms latency \citep{Gookyi2024} and 51.8\,mJ energy per inference \citep{Kouzinopoulos2025}. Most systems rely on post-training quantization and lightweight architectural design, while adaptive offloading strategies have been explored to balance edge and cloud workloads under constrained connectivity \citep{tao2025icrop+}. Despite strong predictive performance, comprehensive reporting of flash, RAM, and energy consumption remains inconsistent.

Several publicly accessible benchmark datasets have played a central role in the development and evaluation of plant disease detection models. Among them, the PlantVillage dataset \citep{hughes2015open}, which contains more than 54,000 leaf images spanning 38 disease categories and 14 crop species, remains the most extensively used benchmark and has been adopted in studies such as \citep{Gookyi2024}, and \citep{Annadata2025}. To address more realistic conditions, the PlantDoc dataset \citep{Singh2020PlantDoc} offers field-acquired images with substantially greater variation in background, illumination, and scene complexity, thereby posing a more challenging evaluation setting. For rice-focused disease detection, the Rice Leaf Disease Dataset used in  \citep{tasci2026dbla} includes field images corresponding to blast, brown spot, and bacterial blight classes. Although these datasets have significantly accelerated model development and benchmarking, an important limitation remains: many of them are derived from controlled laboratory or greenhouse environments, raising concerns about how well they generalize to the heterogeneous and unpredictable conditions of smallholder agricultural fields in real-world deployment scenarios.

\subsection{Precision Irrigation}

Precision irrigation represents a different Edge AI workload class: low-dimensional, sensor-driven inference coupled directly to control and actuation. Compared with vision workloads, its computational demand can be modest, but long-duration operation makes duty cycle, sampling frequency, communication energy, and reliable offline execution especially important. It therefore provides a useful setting for evaluating whether compact regression or neural models can deliver closed-loop decisions within microcontroller-level memory and power budgets.

\begin{figure}[!t]
    \centering
    \includegraphics[width=\linewidth]{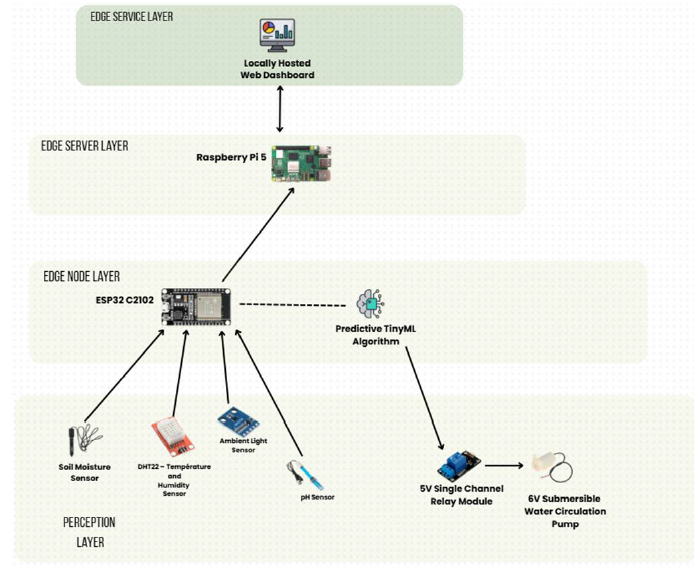}
    \caption{System Prototype: TinyML-Enabled IoT for Sustainable Precision Irrigation (Source: \citep{TinyMLIrrigation}).}
    \label{irrigation}
    \vspace{-0.2in}
\end{figure}

Precision irrigation employs regression models or compact neural networks on ESP32-class microcontrollers \citep{sustainable}, refer Fig.~\ref{irrigation}. Quantized neural networks have achieved highly compact deployments, including a 6.2\,KB model footprint \citep{Hernandez-Hidalgo2026}, enabling fully on-device anomaly detection. Distributed architectures using LoRa communication facilitate coordinated sensing and actuation across spatially dispersed nodes \citep{Thirumalaiah2025}. Although performance metrics such as MAPE and reported water savings demonstrate operational feasibility, systematic energy profiling and long-term deployment evaluation remain limited.

Benchmark datasets for precision irrigation remain comparatively limited, 
and standardized open-access resources for training and evaluating 
irrigation scheduling models are still emerging. One important resource is the OpenET evapotranspiration platform, which provides satellite-derived estimates of crop water demand at field scale with 30~m spatial resolution and can be used as reference data for irrigation scheduling models \citep{Melton2022,volk2024assessing}. The International Soil Moisture Network (ISMN) \citep{dorigo2021international} is another valuable source, offering globally distributed multi-depth soil moisture measurements from more than 2,800 stations, making it suitable for training and validating lightweight regression models. For regions where field-level datasets are limited, regional agrometeorological reanalysis products such as the Indian Monsoon Data Assimilation and Analysis (IMDAA) \citep{rani2021imdaa} provide high-temporal-resolution gridded environmental variables, including temperature, humidity, precipitation, and soil moisture. In addition, \citep{Hernandez-Hidalgo2026} utilizes NDVI anomaly data derived from satellite imagery for model development, representing one of the relatively few studies that explicitly reports the provenance of input data in the context of irrigation-focused edge deployment.

\subsection{Soil Monitoring and Nutrient Analysis}

Soil monitoring and nutrient analysis are dominated by multivariate sensing, periodic inference, and long-duration unattended operation. The systems challenge is less about high-throughput computation and more about integrating heterogeneous sensors, minimizing sampling and communication energy, managing sensor drift, and executing compact models under strict power budgets. This workload therefore exposes the value of cross-layer techniques such as DVFS, event-driven sampling, and scheduling in addition to model-level optimization.

Soil Monitoring and Nutrient Analysis refers to the systematic measurement and evaluation of soil properties to assess soil health, fertility status, and crop suitability. Integration of TinyML models with dynamic voltage and frequency scaling (DVFS) and scheduling mechanisms has demonstrated measurable energy savings, including an 8.5\% reduction in consumption \citep{Bhattacharya2024}. These implementations illustrate the importance of cross-layer optimization linking sensing, computation, and power management in long-duration agricultural deployments. 

Standardized benchmark datasets for soil property estimation and nutrient management are still limited in both scope and geographic representativeness when compared with the wide diversity of agricultural conditions worldwide. Among the available resources, the LUCAS (Land Use/Cover Area frame Survey) topsoil database, maintained by the European Soil Data Centre, is one of the largest harmonized open-access soil datasets in Europe. It includes nearly 45,000 geo-referenced topsoil samples collected across multiple survey cycles from 2009 to 2022, with standardized measurements covering pH, organic carbon, NPK, cation exchange capacity, particle size distribution, and, from 2018 onward, soil biodiversity and pesticide residues \citep{Orgiazzi2022,Fernandez2022}. Owing to its consistency and scale, LUCAS has been widely used for continental-scale digital soil mapping through machine learning approaches such as Random Forest and pedotransfer-function-based modeling \citep{Chen2024LUCAS}. At the global level, the World Soil Information Service (WoSIS), curated by International Soil Reference and Information Centre (ISRIC), provides standardized and quality-controlled soil profile data from more than 228,000 geo-referenced sites across 174 countries \citep{Batjes2024}. In India, the Indian Council of Agricultural Research (ICAR) Soil Health Card scheme has also produced a large volume of geo-referenced soil nutrient records, with more than 23.5 crore cards issued by 2023; however, its utility for machine learning research is still limited by challenges in data standardization, accessibility, and interoperability. Bhattacharya and Pandey \citep{Bhattacharya2024} incorporate sensor-level soil observations within a blockchain-enabled monitoring pipeline, but the provenance of the training data is not publicly reported. This lack of transparency is a common reproducibility issue in soil monitoring research and highlights the urgent need for open, standardized soil datasets that are better suited for developing and benchmarking edge-deployable models.

\subsection{Crop Recommendation}

A crop recommendation system is essentially a decision-making tool for farmers. It helps them select the best crops to grow based on factors such as soil type, climate, available resources and market demand. This field remains comparatively underexplored within TinyML-focused deployments. Classical machine learning models, including Random Forest implementations on ultra-low-cost microcontrollers such as ATMega328P \citep{Baishya2025}, have demonstrated substantial model size reduction (up to 99\%). However, most systems operate with static inference pipelines and lack adaptive learning, distributed coordination, or communication-aware scaling mechanisms.

The Kaggle Crop Recommendation Dataset, which includes soil nutrient attributes (N, P, K), temperature, humidity, pH, and rainfall information across 22 crop categories, has become the most commonly used benchmark for this task \citep{Dey2024} and is also adopted by \citep{Baishya2025}. More recently, the Ethiopian Crop Recommendation Dataset has expanded the scope of available resources by combining geo-referenced soil properties, such as pH, electrical conductivity, and macro- and micronutrient levels, with NASA-derived seasonal climate variables for cereal crops, thereby offering a multimodal dataset rooted in Sub-Saharan African agro-ecological conditions \citep{Demisse2024}. Similarly, \citep{Munir2025} reports the use of an enriched soil fertility dataset containing macro-nutrients, micronutrients, and soil physical characteristics for machine-learning-based crop suitability prediction in South Asian settings. Although these datasets have improved the availability of training data for crop recommendation research, most remain geographically limited and do not adequately reflect important real-world factors such as changing market demand, fluctuations in input costs, and government policy influences, all of which play a significant role in crop selection across diverse agro-ecological environments \citep{Islam2023, Jha2025}.

\subsection{UAV-Assisted and Distributed Intelligence}

UAV-assisted agriculture represents the mobile and distributed end of the Edge AI continuum. These workloads combine high-volume image sensing with strict payload, battery, compute, and communication constraints, and therefore require explicit decisions about what should be processed onboard, at a nearby edge node, or in the cloud. UAV systems are consequently useful for studying adaptive offloading, collaborative inference, split learning, and communication-aware workload partitioning under a moving energy budget.

\begin{figure}[h]
    \centering
    \includegraphics[width=\linewidth]{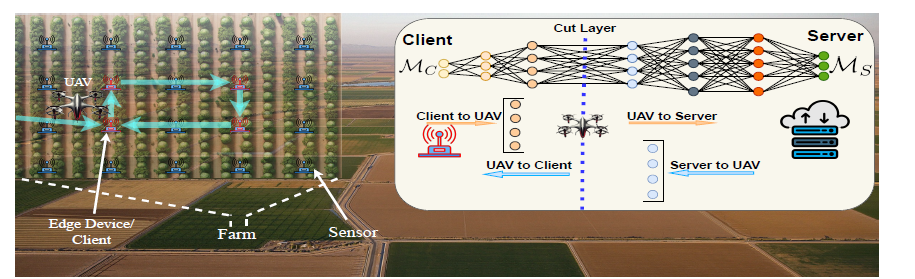}
    \caption{The left sub-figure depicts a single farm with an example of the UAV trajectory for exchanging data with the cluster head devices
(red-colored sensors). The right sub-figure illustrates the communication between a cluster head (client/edge device), a UAV, and a server.(Source: \citep{Soltani2025}).}
    \label{drone}
    \vspace{-0.2in}
\end{figure}

In UAV-Assisted and Distributed Intelligence for precision farming, unmanned aerial vehicles (UAVs or drones) are used as mobile sensing platforms that collect high-resolution aerial data and collaborate with ground-based edge devices to enable localized, intelligent decision-making. UAV-assisted precision agriculture introduces additional constraints related to mobility, payload capacity, and onboard energy availability. Lightweight CNNs adapted through transfer learning have been deployed for drone-based inference to improve operational efficiency \citep{Hayajneh2024}. Distributed learning approaches, including split learning, aim to reduce centralized training overhead and have reported up to an 86\% energy reduction in distributed settings \citep{Soltani2025}; see Fig.~\ref{drone}. Nevertheless, these strategies introduce communication dependencies that may limit applicability in connectivity-constrained rural environments.

Several UAV-based agricultural datasets have been introduced to facilitate aerial crop analysis. Among the most prominent, the Agriculture-Vision dataset offers large-scale aerial farmland imagery consisting of about 94,986 RGB-NIR images captured over U.S. farmlands at a spatial resolution of 10~cm/pixel, together with pixel-level semantic segmentation labels for eight agricultural conditions, including weeds, nutrient deficiency, and drydown \citep{Chiu2020}. The Extended Agriculture-Vision dataset further expands this resource by adding 3,600 full-field images and benchmarks for self-supervised pre-training \citep{Wu2023}. For weed-focused applications, the CoFly-WeedDB dataset contains UAV-acquired RGB images captured at an altitude of 5~m over cotton fields in Greece, annotated for three major weed species \citep{Shahi2023}. Similarly, the DRONEWEED dataset includes more than 67,000 labeled UAV images collected at 11~m altitude across maize and tomato fields in Spain, covering ten weed species at two different phenological stages \citep{Mesias2025}. For crop health assessment, Jadhav et al. (2025) present an Indian UAV and leaf image dataset for soybean, spanning multiple growth stages and including annotations for both diseases and pests across two growing seasons \citep{Jadhav2025}. Hayajneh et al. \citep{Hayajneh2024} also demonstrate the use of transfer learning from ImageNet-pretrained models to adapt learned visual features for aerial crop imagery. Despite these contributions, the literature still lacks standardized UAV-specific benchmark datasets with consistent resolution, flight altitude, and ground-truth annotation protocols designed specifically for evaluating lightweight, edge-deployable models, which remains a major challenge for resource-constrained agricultural deployment scenarios \citep{Anam2024,Zhu2024}.

\begin{table*}[!t]
\centering
\caption{System-level intelligence-placement patterns.}
\label{tab:intelligence_placement}
\vspace{-0.1in}
\setlength{\tabcolsep}{4pt}
\renewcommand{\arraystretch}{1.3}

\resizebox{\textwidth}{!}{
\begin{tabular}{lllll}
\toprule
\textbf{References} & \textbf{Intelligence Distribution} & \textbf{Edge--Cloud Dependency} & \textbf{Communication Layer} & \textbf{Inference / Training Tier} \\
\midrule

\citep{Samanta2025}
& Fully On-Device
& None
& Not Specified
& On-Device Inference / Cloud Training \\

\citep{tao2025icrop+} 
& Edge-Assisted 
& Optional 
& LoRa + IoT Network 
& Hybrid Inference / Cloud Training \\

\citep{Thirumalaiah2025} 
& Edge-Coordinated 
& Optional 
& LoRa 
& Edge Inference / Cloud Training \\

\citep{Soltani2025} 
& Distributed Collaborative 
& Required 
& Wireless Edge Link 
& Edge Inference / Distributed Training \\

\citep{Bhattacharya2024} 
& Edge-Assisted 
& Optional 
& IoT Network 
& Node Inference / Cloud Training \\

\citep{Annadata2025} 
& Fully On-Device 
& None 
& UAV Internal Control 
& On-Device Inference / Cloud Training \\

\citep{Madiwal2025} 
& Edge-Assisted 
& Optional 
& WiFi / IoT 
& Edge Inference / Cloud Training \\

\citep{sustainable} 
& Edge-Assisted 
& Optional 
& IoT Network 
& Edge Inference / Cloud Training \\

\citep{Hernandez-Hidalgo2026} 
& Fully On-Device 
& None 
& Not Specified 
& On-Device Inference / Cloud Training \\

\bottomrule
\end{tabular}
}
\end{table*}

\vspace{-0.15in}
\section{System-Level Architectural and Deployment Patterns }
\label{sec:system_patterns}
\vspace{-0.1in}

This section directly addresses \textbf{RQ3} by examining where inference, coordination, and learning responsibilities are placed across the device--edge--cloud continuum. Table~\ref{tab:intelligence_placement} organizes intelligence placement in Edge AI and TinyML-based precision agriculture into three dominant architectural paradigms: \textit{fully on-device}, \textit{edge-assisted}, and \textit{distributed collaborative} systems. These configurations differ in computational locality, communication dependency, and inference--training separation, directly shaping latency, energy efficiency, and operational resilience in resource-constrained environments.

\subsection{Fully On-Device Architectures}

Fully on-device systems perform inference entirely on embedded or microcontroller-class hardware without runtime cloud dependency. This specification is illustrated in Table~\ref{tab:intelligence_placement}. \citep{Hernandez-Hidalgo2026} implements an 8-bit quantized neural network for irrigation anomaly detection with a compact 6.2\,KB model footprint, thus establishing strict memory feasibility for microcontroller-class deployment. Similarly, \citep{Annadata2025} deploys CNN-based inference using ESP-EYE hardware for UAV-assisted crop monitoring, attaining 90\% classification accuracy and embedding perception within the drone’s control loop. \citep{Samanta2025} deployed CNN model occupying 103.9 KB RAM size on ESP32-CAM for tomato leaf disease detection on fully on-device inference mode.

These architectures mitigate communication latency and bandwidth burden, which allows robust field operation. However, localized inference does not necessarily eliminate end-to-end cloud dependency for training and model updates, exposing an inevitable inference-training asymmetry.

\subsection{Edge-Assisted Architectures}
\vspace{-0.01in}
Edge-assisted systems inference locally; optional cloud interaction for both training and coordination is retained. This trend is repeated in \citep{tao2025icrop+}, \citep{Thirumalaiah2025}, \citep{Bhattacharya2024}, \citep{Madiwal2025}, and \citep{sustainable}, exploiting LoRa, WiFi, or IoT networks for distributed sensing.

\cite{tao2025icrop+} introduces adaptive offloading over LoRa to dynamically balance edge and cloud workloads. \cite{Bhattacharya2024} combine TinyML inference with DVFS-based power management, achieving an 8.5\% energy reduction, while \cite{Thirumalaiah2025} demonstrate LoRa-coordinated irrigation control with measurable water savings. 

While these architectures offer scalability and centralized manageability, optional edge–cloud dependency brings sensitivity to network instability. However, latency-critical inference is still localized.

\subsection{Distributed Collaborative Architectures}

Distributed collaborative systems share intelligence across multiple nodes and partition learning responsibilities. \citep{Soltani2025} realize split learning via edge inference and distributed training, yielding up to 86\% energy reduction in contrast to centralized techniques. Such frameworks need constant communication to synchronize models in contrast to edge-assisted systems. Although good for efficient training and distributed computation, compulsory connectivity can be a constraint for rural deployments that have intermittent infrastructure. 

\subsection{Implications}
Across the configurations in Table~\ref{tab:intelligence_placement}, a consistent systems-level pattern emerges: inference is increasingly localized, whereas training and lifecycle management remain predominantly centralized. Quantitative findings such as the 6.2~KB model in \citep{Hernandez-Hidalgo2026}, 92\% classification accuracy in \citep{Samanta2025}, 8.5\% energy reduction in \citep{Bhattacharya2024}, 90\% UAV accuracy in \citep{Annadata2025}, and 86\% distributed energy savings in \citep{Soltani2025} show that localized execution is feasible across very different resource budgets. The more consequential architectural observation, however, is the separation of fast local inference from centralized model development and updating.

This inference--training asymmetry offers low-latency autonomy while retaining the computational capacity of centralized training, but it also creates a lifecycle dependency: model refresh, adaptation, versioning, and redistribution remain external to the endpoint. For resource-constrained deployments, the practical design problem is therefore not simply ``edge versus cloud,'' but how inference, adaptation, coordination, and model management should be partitioned across endpoint, gateway, and cloud tiers under communication and energy constraints.

\begin{table*}[!t]
\centering
\caption{Optimization strategies identified.}
\label{tab:optimization_strategies}
\vspace{-0.1in}
\setlength{\tabcolsep}{4pt}
\renewcommand{\arraystretch}{1.3}

\resizebox{\textwidth}{!}{
\begin{tabular}{lllllllll}
\toprule
\textbf{References} & \textbf{Quant.} & \textbf{Pruning} & \textbf{Model Comp.} & \textbf{Transfer Learning} & \textbf{Split Learning} & \textbf{DVFS} & \textbf{Adaptive Offloading} & \textbf{Lightweight Arch.} \\
\midrule

\citep{Samanta2025} & Yes & -- & -- & -- & -- & -- & -- & Yes \\
\citep{Koli2025} & Yes & -- & Yes & -- & -- & -- & -- & Yes \\
\citep{Gookyi2024} & -- & -- & -- & -- & -- & -- & -- & Yes \\
\citep{tao2025icrop+} & -- & -- & -- & -- & -- & -- & Yes & Yes \\
\citep{Kouzinopoulos2025} & Yes & Yes & -- & -- & -- & -- & -- & Yes \\
\citep{Madiwal2025} & Yes & -- & -- & -- & -- & -- & -- & Yes \\
\citep{sustainable} & -- & -- & -- & -- & -- & -- & -- & Yes \\
\citep{Hernandez-Hidalgo2026} & Yes & -- & Yes & -- & -- & -- & -- & Yes \\
\citep{Thirumalaiah2025} & -- & -- & -- & -- & -- & -- & -- & Yes \\
\citep{Bhattacharya2024} & -- & -- & -- & -- & -- & Yes & -- & -- \\
\citep{Baishya2025} & -- & -- & Yes & -- & -- & -- & -- & Yes \\
\citep{Soltani2025} & -- & -- & -- & -- & Yes & -- & -- & -- \\
\citep{Hayajneh2024} & Yes & -- & -- & Yes & -- & -- & -- & Yes \\

\bottomrule
\end{tabular}
}
\vspace{-0.05in}
\end{table*}

\begin{table*}[!t]
\centering
\caption{Hardware-resource and energy profiling reported.}
\label{tab:energy_resource}
\vspace{-0.1in}
\setlength{\tabcolsep}{7pt}
\renewcommand{\arraystretch}{1.2}

\resizebox{\textwidth}{!}{
\begin{tabular}{llllll}
\toprule
\textbf{References} & \textbf{Flash / ROM} & \textbf{RAM} & \textbf{Model Size} & \textbf{Latency} & \textbf{Energy / Inference or Reduction Rate} \\
\midrule
\citep{Gookyi2024} & 344.7 KB & 726.6 KB & CNN-MDD & 7.6 ms & -- \\
\citep{Kouzinopoulos2025} & 850.97 KB & 677.30 KB & Pruned INT8 YOLOv8n & -- & 51.8 mJ (Device inference) \\
\citep{Hernandez-Hidalgo2026} & -- & -- & 6.2 KB Model & -- & -- \\
\citep{Baishya2025} & -- & -- & 99\% Size Reduction & -- & -- \\
\citep{Bhattacharya2024} & -- & -- & -- & -- & 8.5\% Energy $\downarrow$ (DVFS/system operation)\\
\citep{Soltani2025} & -- & -- & -- & -- & upto 86\% Energy $\downarrow$ (Distributed/split-learning setting) \\
\citep{Hayajneh2024} & -- & -- & TL-based CNN & -- & -- \\
\citep{Samanta2025} & 323 KB & 103.9 KB & Quantized CNN & 850 ms & -- \\
\citep{Madiwal2025} & -- & -- & Quantized MobileNet & -- & -- \\
\citep{tasci2026dbla} & -- & -- & DBLA-MobileNetV2 (FP16) & -- & -- \\
\bottomrule
\end{tabular}
}
\vspace{-0.2in}
\end{table*}

\section{Optimization Strategies and Resource Profiling}
\label{sec:optimization_profiling}
\vspace{-0.1in}
This section jointly addresses \textbf{RQ2} and \textbf{RQ4}: it first examines how deployments reduce or redistribute computational cost, and then evaluates whether the resource measurements needed to verify those deployment claims are reported consistently. Optimization in resource-constrained Edge AI must be evaluated against the physical deployment budget rather than predictive accuracy or parameter count alone. Tables~\ref{tab:optimization_strategies} and \ref{tab:energy_resource} are interpreted across three interacting levels: model-level compression and architecture design, compiler/runtime realization, and system-level sensing-communication-power management. A central contribution of this study is the accompanying resource-profiling analysis, which asks whether the measurements required to judge actual deployability, such as flash, peak RAM, computational workload, latency, power, and energy per inference, are reported consistently.

\subsection{Model-Level Optimization Patterns}
Among the 13 works summarized in Table~\ref{tab:optimization_strategies}, 6 report quantization. Specifically, \citep{Samanta2025}, \citep{Koli2025}, \citep{Kouzinopoulos2025}, \citep{Madiwal2025}, \citep{Hernandez-Hidalgo2026}, and \citep{Hayajneh2024} employ INT8 or reduced-precision execution to reduce model footprint and computational overhead. The evidence therefore indicates that precision reduction is currently the most mature optimization mechanism within this subset.

The reported deployments also show why model size alone is insufficient. \citep{Hernandez-Hidalgo2026} report an 8-bit neural network with a 6.2~KB model footprint for microcontroller-level anomaly detection, whereas \citep{Kouzinopoulos2025} deploy INT8 YOLOv8n on STM32U5 hardware with 850.97~KB flash, 677.30~KB RAM, and 51.8~mJ per inference. \citep{Gookyi2024} report 7.6~ms inference with 344.7~KB flash and 726.6~KB RAM. These systems occupy very different points in the memory--compute design space despite all being described as edge or TinyML deployments.

Structured compression is considerably less common. Explicit structured pruning appears in \citep{Kouzinopoulos2025}, while classical model compression in \citep{Baishya2025} reports up to 99\% model-size reduction for crop recommendation on ATMega328P. The limited use of structured pruning, multi-objective compression, and hardware-aware neural architecture search suggests that current practice still favors precision scaling and manually selected lightweight architectures over systematic structural reconfiguration.

\begin{figure}[!t]
    \centering
\includegraphics[width=\linewidth]{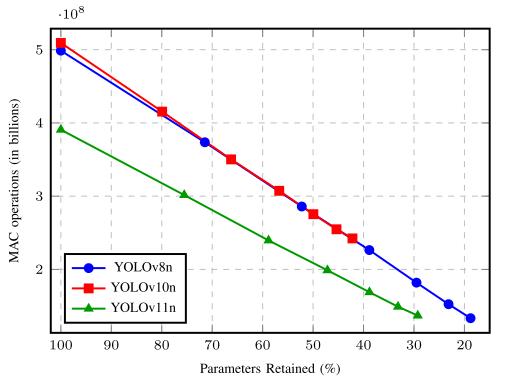}
    \caption{Effect of pruning on inference complexity for YOLOv8n, YOLOv10n, and YOLOv11n (Source: \citep{Kouzinopoulos2025}).}
    \label{pruning}
    \vspace{-0.2in}
\end{figure}

\subsection{Compiler- and Runtime-Level Optimization}
A compressed model does not automatically translate into an efficient executable. Realized latency, memory use, and energy depend on whether the deployment runtime supports the selected numerical precision and operators, whether optimized kernels are available, and how intermediate tensors are allocated. The toolchains discussed in Section~\ref{subsec:tinyml_toolchains} therefore form a distinct optimization layer through graph rewriting, operator fusion, static memory planning, kernel selection, and target-specific code generation. Future comparisons should identify the runtime and conversion path used for every reported hardware measurement so that model-level gains can be separated from compiler- or kernel-level gains.

\subsection{Distributed and System-Level Optimization}
Optimization can also occur outside the model. \citep{Soltani2025} use split learning and report up to 86\% lower energy consumption than centralized training in their evaluated setting, while \citep{tao2025icrop+} explores adaptive offloading over LoRa-based networks. These strategies distribute computation across devices and infrastructure, but they also introduce communication costs and dependence on link availability. Their benefit therefore needs to be assessed through an end-to-end budget that includes data transfer, waiting time, retransmission, and radio energy rather than compute cost alone.

\subsection{Hardware-Aware and Cross-Layer Co-Design}
Explicit hardware--algorithm co-optimization remains comparatively rare. \cite{Bhattacharya2024} combines TinyML inference with dynamic voltage and frequency scaling (DVFS) and reports an 8.5\% energy reduction, illustrating that power management can complement model optimization. More broadly, sensing rate, input resolution, preprocessing, inference frequency, clock scaling, and communication policy jointly determine the resource envelope. Hardware-aware NAS \citep{saha2025tinytnas}, structured multi-objective pruning \citep{saha2025gencprunex}, and efficiency-aware data acquisition \citep{saha2025efficiency} point toward this broader co-design space, but these strategies remain underrepresented in the analyzed farming deployments.

\subsection{Resource Profiling as a Deployment-Readiness Criterion}
Table~\ref{tab:energy_resource} exposes one of the most consequential gaps in the current evidence base: hardware-level reporting is fragmented and non-uniform. Individual studies provide valuable measurements such as 51.8~mJ per inference \citep{Kouzinopoulos2025}, 7.6~ms latency \citep{Gookyi2024}, or substantial model-size reduction \citep{Baishya2025}, but the measurements are rarely reported as a consistent set. Peak RAM, flash usage, MAC or operation count, measured latency, average or peak power, and energy per inference are frequently absent.

These metrics are not interchangeable. Model-file size does not reveal peak activation memory; MAC count does not necessarily predict latency on a specific processor and runtime; latency alone does not establish battery feasibility; and quantization does not guarantee energy reduction unless the target kernels efficiently execute lower-precision arithmetic. For MCU-class systems, a deployment-ready evaluation should therefore report, at minimum, the target device and clock configuration, numerical precision, program or flash footprint, peak SRAM or tensor-arena requirement, operation count where applicable, measured inference latency, and energy or power under a stated measurement protocol. Sensing and communication overhead should be reported separately when they materially contribute to the end-to-end budget.

Accordingly, resource profiling is treated here as a primary criterion for deployment readiness rather than as an implementation detail. Without a reproducible resource profile, it is difficult to compare two purportedly lightweight systems or to determine whether an implementation can transfer to a device with a different memory hierarchy, runtime, or power envelope.

\subsection{Implications}
The combined evidence suggests a transition from single-technique model compression toward multi-level deployment optimization. Quantization and lightweight architectures are currently the dominant mechanisms, while structured pruning, hardware-aware architecture search, compiler-aware execution, adaptive sensing, and communication-aware workload placement remain less mature. The resulting design objective is therefore multi-dimensional: model precision and structure should be co-optimized with runtime support, peak memory, sensing rate, communication policy, and device-level power management. A system should not be considered deployment-ready solely because the model fits in storage or achieves high predictive accuracy; it should demonstrate that the complete inference path satisfies the target platform's memory, latency, energy, and connectivity constraints.

\section{Layered Deployment Perspective for Resource-Constrained Edge AI and TinyML}
\label{sec:deployment_perspective}

\begin{figure*}[t]
    \centering
    \includegraphics[width=\linewidth]{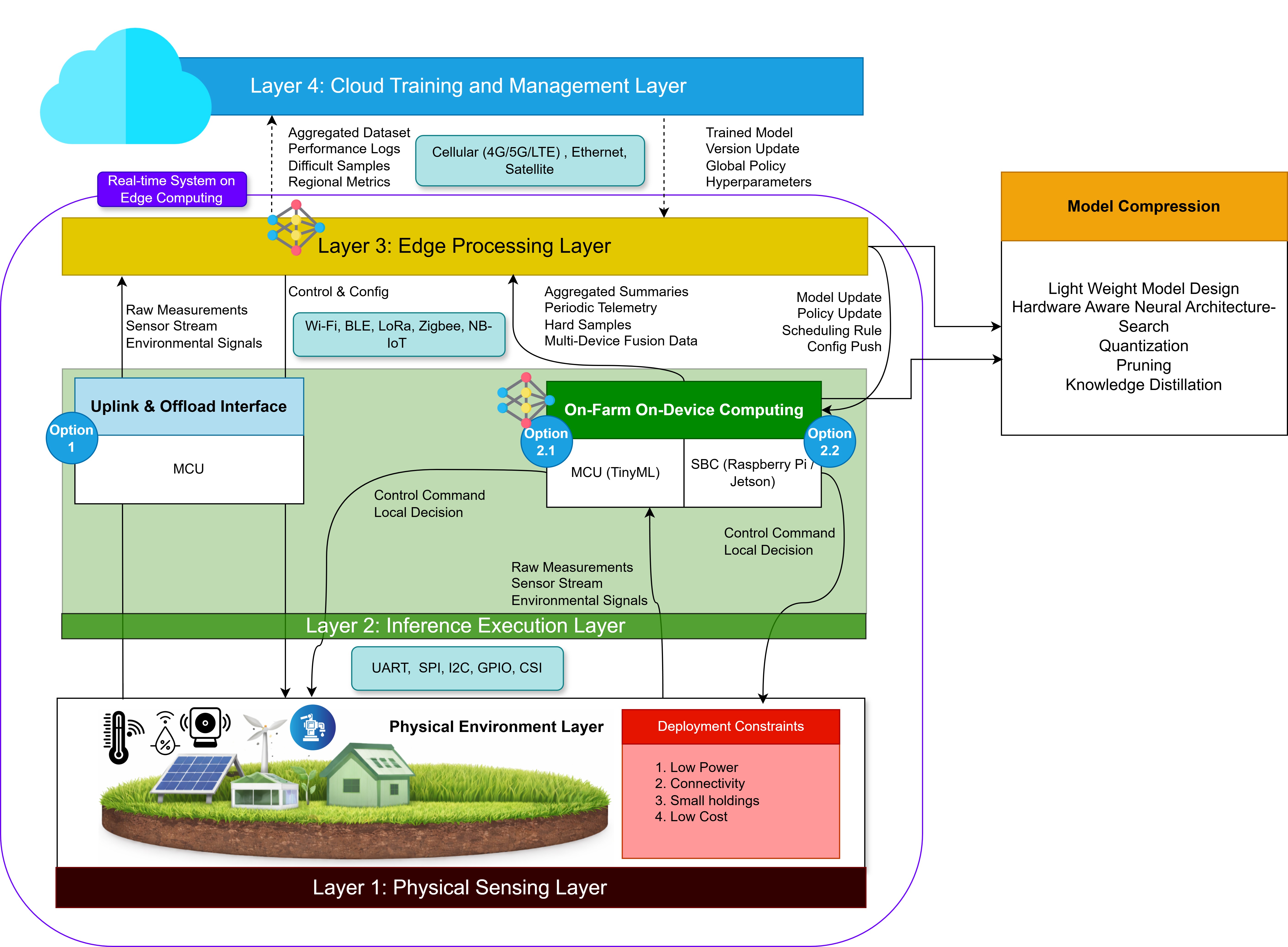}
    \caption{Deployment architecture for resource-constrained Edge intelligence in precision farming.}
    \label{fig:edge-ai-architecture}
\end{figure*}

Based on the cross-question synthesis above, we propose a \textit{layered deployment perspective for resource-constrained Edge AI and TinyML}. The purpose of this perspective is not to define an agriculture-specific AI stack, but to make the placement of sensing, inference, near-edge processing, and centralized training explicit under heterogeneous resource budgets. Figure~\ref{fig:edge-ai-architecture} instantiates this generic perspective in precision farming, where environmental sensors, imaging platforms, actuators, gateways, and cloud services provide concrete examples of the four layers. The same deployment logic can be applied to other distributed cyber-physical settings in which endpoints operate under constrained memory, power, connectivity, or maintenance conditions.

The architecture should be interpreted as a resource-aware decision framework rather than as a requirement that every deployment use all four layers. A workload may remain entirely on-device, may offload selected operations to a local edge node, or may use the cloud only for training and lifecycle management. The placement decision is governed by latency, privacy, available memory and compute, energy budget, communication reliability, and the cost of moving data. In this way, the perspective operationalizes the findings of RQ1--RQ4: hardware and toolchain capability determine feasible execution targets, optimization determines whether a workload fits those targets, intelligence placement determines communication dependence, and resource profiling verifies whether the chosen deployment is sustainable.

\subsection{Layer 1: Physical Sensing Layer}
The Physical Sensing Layer interfaces the computing system with the environment. Its primary responsibilities are data acquisition, actuator interfacing, low-level signal handling, and power-aware peripheral control. The precision-farming instantiation in Fig.~\ref{fig:edge-ai-architecture} includes temperature, humidity, wind, soil-moisture, and imaging sensors, including UAV-mounted cameras, together with actuators such as pumps, motors, relays, and fans. These devices are application-specific examples of a more general edge-system function: converting physical phenomena into data streams and applying local control actions.

From an Edge AI or TinyML perspective, this layer is important because sensing decisions establish the resource demand seen by every upper layer. Sampling frequency, image resolution, sensor duty cycle, buffering, and local preprocessing directly affect memory occupancy, computation, communication volume, and energy consumption. Hardware-level interfaces such as \textit{I2C, SPI, UART, ADC, PWM, GPIO,} and \textit{USB} connect the sensing and actuation elements to the execution hardware. Consequently, sensing should be treated as part of the deployment budget rather than as a fixed front end external to AI optimization.

\subsection{Layer 2: Inference Execution Layer}
The Inference Execution Layer is the principal location for latency-sensitive analytics and represents the point at which the system decides whether computation should remain at the endpoint or be transferred upward. It therefore supports two complementary modes: \textit{(i) computational offloading} and \textit{(ii) on-device computing}. The appropriate mode depends on the workload's memory and compute requirements, the capabilities of the target processor and runtime, the energy cost of communication, and the availability of a reliable link.

\subsubsection{Computational Offloading}
In the computational-offloading mode, the endpoint performs sensing and lightweight preprocessing but forwards data, features, or selected events to the Edge Processing Layer for inference. The resulting decision is then returned to the local controller or actuator. Offloading is appropriate when the endpoint lacks sufficient memory, supported operators, accelerator capability, or latency headroom to execute the complete model locally. However, it exchanges local computational cost for communication cost and therefore becomes less attractive when radio energy, bandwidth, or link availability dominates the deployment budget. For this reason, offloading should be selected using end-to-end measurements rather than model complexity alone.

\subsubsection{On-Device Computing}
In the on-device mode, inference is executed at or immediately adjacent to the sensing source. This minimizes round-trip latency, reduces transmission of raw data, and can preserve operation during network outages. Two practical execution regimes shown in the architecture are TinyML-based inference and more capable SBC/accelerator-based Edge AI.



Across the on-device regimes, model optimization and runtime realization remain inseparable. Lightweight architectures, hardware-aware neural architecture search, knowledge distillation, pruning, quantization, optimized kernels, and target-specific compilation can all change the feasible execution envelope. After inference, the endpoint may trigger an actuator locally and transmit only the prediction, compressed metadata, or selected telemetry to the Edge Processing Layer for coordination and historical logging. Depending on range, bandwidth, and power constraints, this communication may use \textit{LoRa, LoRaWAN, NB-IoT, Wi-Fi, BLE,} or \textit{Zigbee}. In the farming instantiation, this enables local control even when continuous cloud access is unavailable.

\subsection{Layer 3: Edge Processing Layer}
The Edge Processing Layer provides near-edge computation that is more capable than the endpoint but remains geographically and logically closer to the data source than the cloud. It can aggregate multiple devices, perform multi-sensor fusion, execute models that do not fit on constrained endpoints, coordinate local control, buffer data during network outages, and support collaborative or partitioned inference. It is also the natural location for workload orchestration: a gateway can decide which events remain local, which require a larger model, and which summaries should be forwarded to the cloud.

This layer is particularly important when a deployment contains many resource-constrained nodes. Rather than equipping every endpoint with high-capability hardware, shared edge processing can amortize computation and storage while preserving lower latency and lower upstream bandwidth than a cloud-only design. In Fig.~\ref{fig:edge-ai-architecture}, the farm-level edge node exemplifies this role, but the same layer can be implemented by an industrial gateway, mobile edge computer, local server, or capable UAV platform. Communication to the cloud may use \textit{cellular networks, satellite links, Wi-Fi,} or \textit{Ethernet}, depending on the deployment context.

\subsection{Layer 4: Cloud Training and Management Layer}
The Cloud Training and Management Layer provides the compute- and storage-intensive functions that are difficult or inefficient to maintain on resource-constrained devices. These include large-scale dataset aggregation, model training and retraining, experiment management, model versioning, configuration management, long-term telemetry storage, fleet-level analytics, and distribution of validated model or parameter updates. In the reverse direction, endpoints and edge nodes can upload selected observations, resource logs, drift indicators, and performance summaries rather than continuously transmitting all raw data.

This layer captures the inference-training asymmetry identified in RQ3. The architecture intentionally keeps time-critical inference near the data source while allowing training and much of the model lifecycle to remain centralized. The design challenge is therefore not to eliminate the cloud, but to minimize unnecessary dependence on it while maintaining safe model updates, reproducibility, and adaptation. Where appropriate, federated, split, or selective-update mechanisms can shift parts of learning toward the edge, but these mechanisms should be evaluated against their communication, storage, and energy overheads.

\subsection{Design Interpretation and Transferability}
The four layers form a continuum of computational capability rather than four mandatory processing stages. A simple TinyML node may use only Layers~1 and~2 during normal operation and contact Layer~4 occasionally for model updates; a camera-heavy workload may use Layers~1--3 for local inference and aggregation; a distributed deployment may use all four. The architecture therefore supports multiple intelligence-placement patterns identified in Table~\ref{tab:intelligence_placement} without prescribing one as universally preferable.

Precision farming remains the concrete application shown in Fig.~\ref{fig:edge-ai-architecture} because it combines heterogeneous sensing, intermittent connectivity, autonomous actuation, and strong cost and energy constraints in a single deployment environment. However, the contribution of the perspective is the underlying Edge AI/TinyML systems logic: match workload complexity to an execution tier, minimize unnecessary data movement, retain latency-sensitive decisions near the endpoint, centralize only the functions that require larger compute or global information, and verify each placement decision through explicit resource profiling. 

\begin{figure*}[!t]
\centering

\begin{subfigure}[t]{0.32\textwidth}
    \centering
    \includegraphics[width=\linewidth]{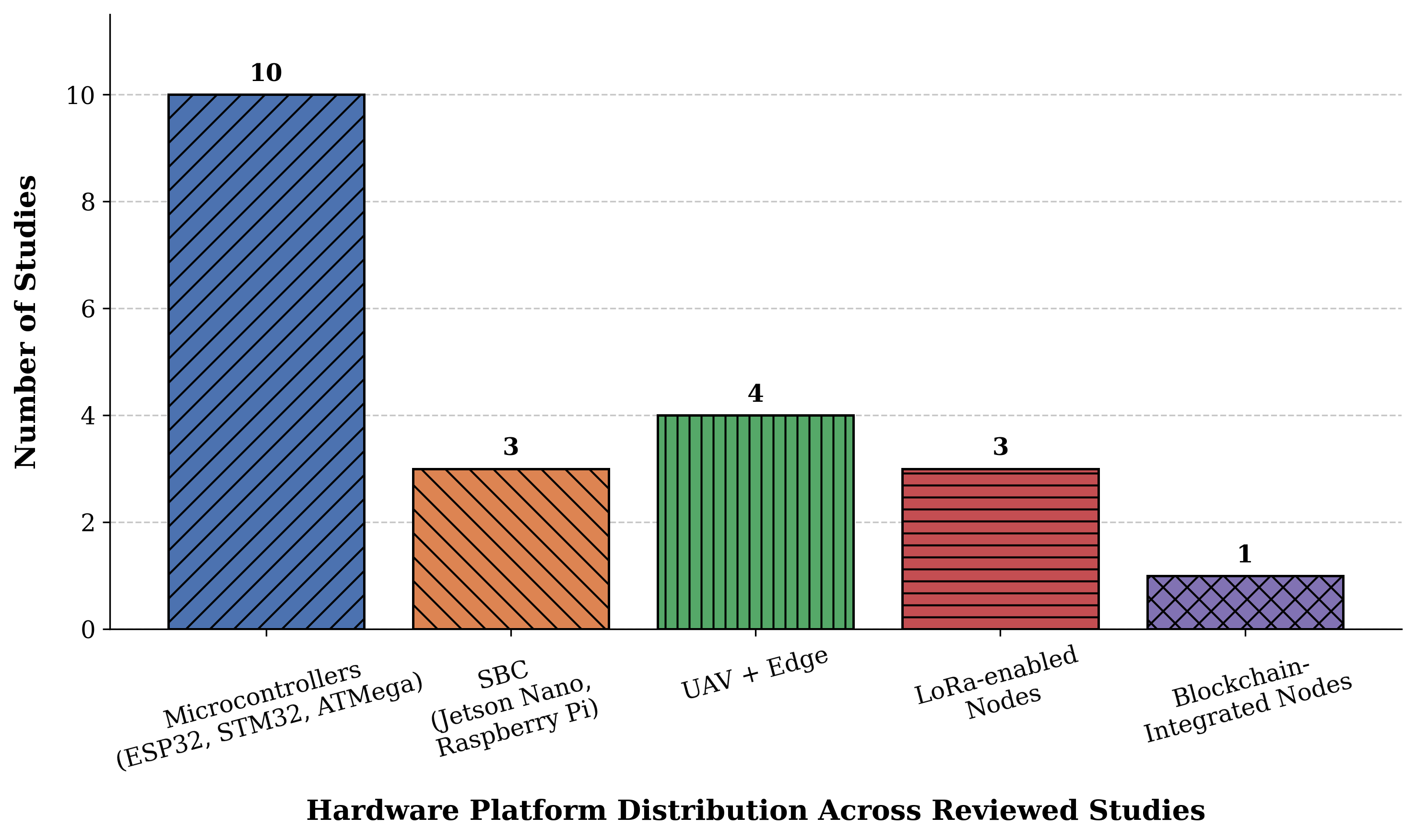}
    \caption{Hardware Platform Distribution}
    \label{fig:hardware_dist}
\end{subfigure}
\hfill
\begin{subfigure}[t]{0.32\textwidth}
    \centering
    \includegraphics[width=\linewidth]{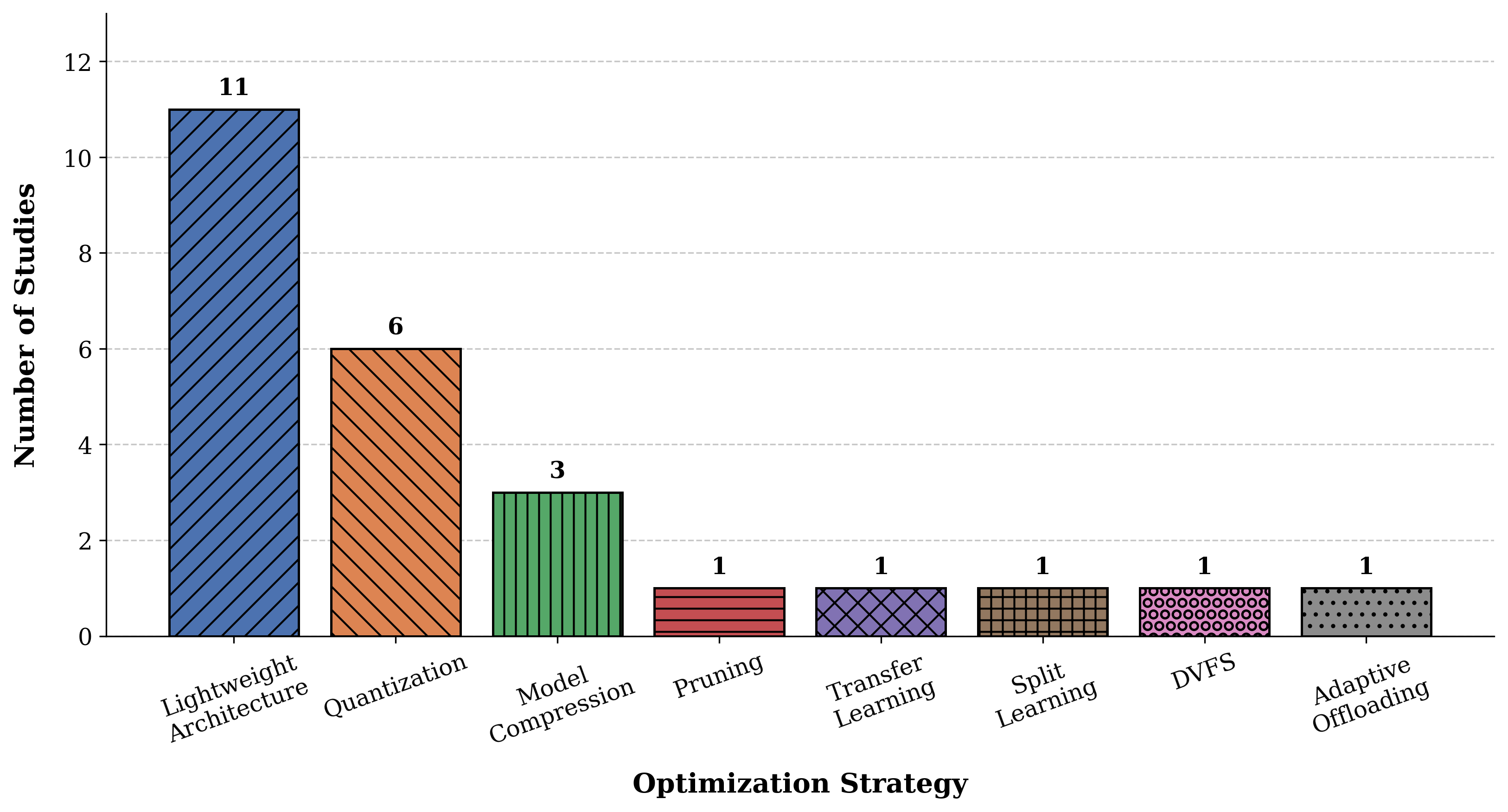}
    \caption{Dominant Optimization Patterns}
    \label{fig:optimization_dist}
\end{subfigure}
\hfill
\begin{subfigure}[t]{0.32\textwidth}
    \centering
    \includegraphics[width=\linewidth]{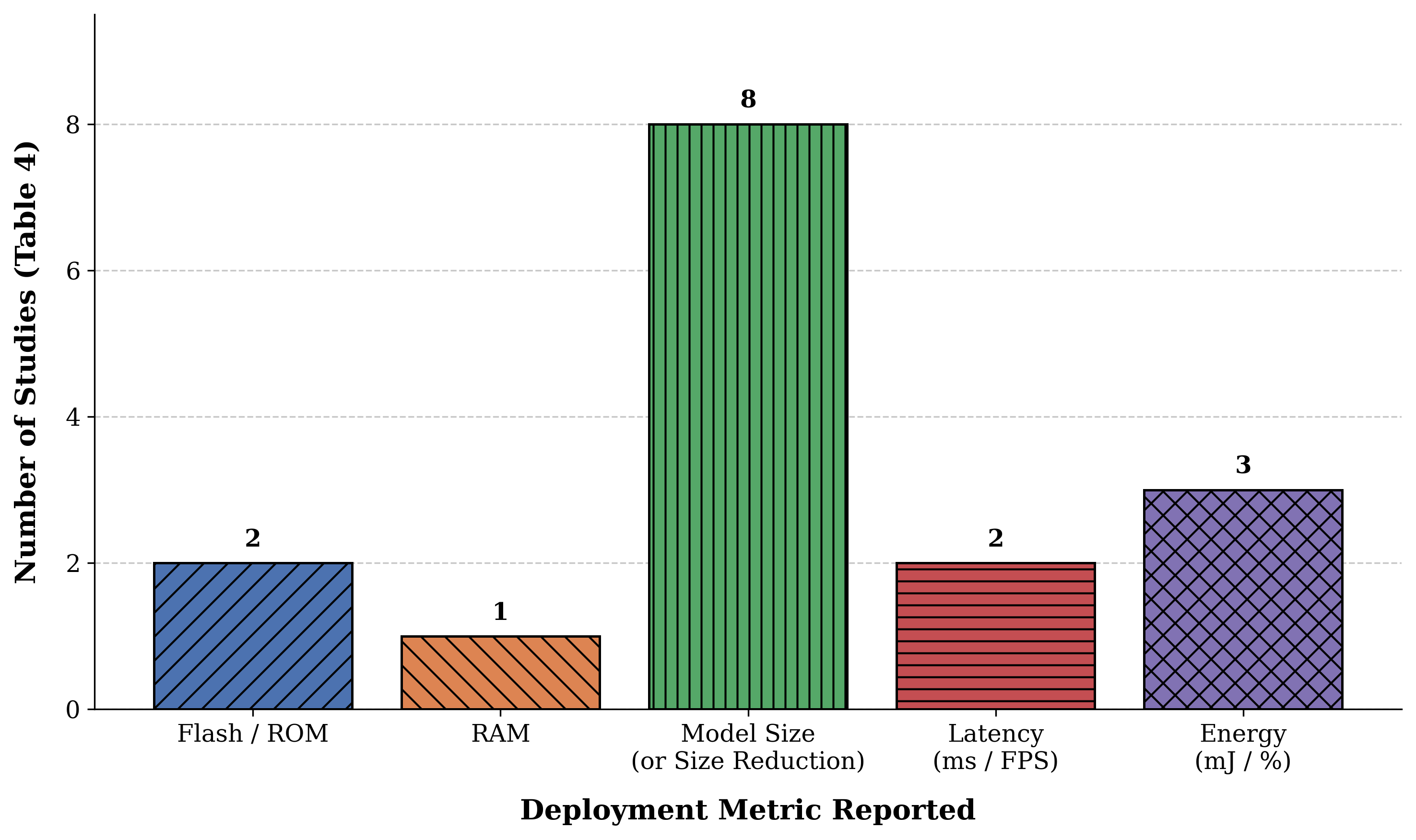}
    \caption{Resource Reporting Gap}
    \label{fig:reporting_gap}
\end{subfigure}
\caption{Deployment-oriented evidence summary: (a) hardware platform distribution, (b) optimization strategies, and (c) reporting frequency of hardware-resource metrics.}
\label{fig:deployment_analysis}
\vspace{-0.2in}
\end{figure*}

\vspace{-0.1in}

\vspace{-0.2in}
\section{Open Challenges and Research Directions}
\vspace{-0.1in}

The next stage of Edge AI and TinyML research in precision farming is fundamentally a systems problem. The literature demonstrates that compact models can run on constrained hardware, but deployment maturity now depends on whether models, runtimes, sensing pipelines, communication policies, and lifecycle mechanisms are designed and evaluated jointly.

\subsection{Standardized Resource-Aware Benchmarking}
The most immediate need is a reproducible deployment benchmark that complements predictive metrics with hardware measurements. Flash or program size, peak SRAM, operation count, measured latency, average or peak power, and energy per inference should be reported together with the target processor, clock configuration, numerical precision, runtime, compiler or toolchain, and measurement method. The current evidence shows that isolated metrics are often available, but complete resource profiles are uncommon. This prevents meaningful cross-device comparison and makes it difficult to determine whether improvements arise from the model architecture, the runtime, the processor, or the measurement setting. Future benchmarks should therefore separate model-level complexity from realized on-device performance and should include sensing and communication overhead whenever these dominate the end-to-end budget.

\subsection{Training--Inference Asymmetry and Adaptive Edge Intelligence}
A recurring architectural pattern across the analyzed systems is the localization of inference while training and model lifecycle management remain centralized. This inference--training asymmetry is attractive because it keeps endpoint execution simple and energy efficient, but it also creates a dependence on cloud-managed retraining, model distribution, and update mechanisms. In dynamic field environments, such separation can limit adaptation to sensor drift, seasonal change, new disease appearances, or site-specific conditions.

Future research should investigate how much adaptation can be moved toward the edge without violating memory and energy constraints. Communication-efficient federated learning, split learning, lightweight continual learning, parameter-efficient update mechanisms, and selective on-device calibration are relevant directions, but their evaluation must include the communication and storage cost of model updates rather than training accuracy alone. The key systems question is not whether all training should move to the endpoint, but how training, adaptation, and inference should be partitioned across endpoint, gateway, and cloud tiers.

\subsection{Cross-Layer Sensing--Compute--Communication Co-Design}
Most deployments optimize the model after data-acquisition choices have already been fixed. This separation can be inefficient because input resolution, sampling frequency, camera duty cycle, sensor polling rate, preprocessing, radio transmission, and wake-sleep scheduling may consume resources comparable to or greater than inference itself. Future TinyML systems should therefore optimize the complete sensing--compute--communication pipeline. Adaptive sampling, event-triggered inference, DVFS, communication-aware offloading, and workload scheduling should be evaluated jointly with quantization, pruning, and architecture search.

Hardware-aware neural architecture search is particularly promising when extended beyond MAC count or model size to measured latency, peak activation memory, energy, and supported operators on the target runtime \citep{saha2025tinytnas}. Similarly, structured pruning and acquisition-rate reduction should be assessed through their device-level consequences rather than sparsity or theoretical complexity alone \citep{saha2025gencprunex,saha2025efficiency}.

\subsection{Compiler, Runtime, and Lifecycle Robustness}
Deployment reliability depends on more than successful model conversion. Operator support, static memory allocation, quantization compatibility, optimized-kernel availability, firmware integration, model versioning, remote update mechanisms, sensor recalibration, and fault recovery determine whether an Edge AI node remains usable over long periods. These factors are especially important for unattended or difficult-to-service deployments. Future work should report the complete software stack and evaluate update and maintenance pathways alongside inference performance.

Long-term sustainability also requires accounting for battery aging, hardware durability, enclosure and environmental protection, maintenance frequency, and replacement cost. These factors connect low-level systems engineering to economic deployability and should become part of multi-season evaluations rather than afterthoughts following laboratory validation.

\subsection{Field Validation and Transferability}
A persistent gap remains between successful execution on a development board and reliable operation in the target environment. Changing illumination, dust, moisture, temperature, sensor drift, heterogeneous backgrounds, and user interaction can alter both predictive behavior and system resource consumption. The evidence in Table~\ref{tab:application_summary} includes physical deployments alongside simulation-based and conceptual studies, underscoring the need to distinguish device execution from sustained field validation.

Future studies should therefore combine technical and application-level evaluation. At the systems level, this includes memory headroom, latency distribution, energy over realistic duty cycles, communication outages, and failure recovery. At the application level, it includes whether the deployed system produces useful outcomes such as earlier detection, water savings, or reduced intervention frequency. Such multi-level validation is necessary to determine whether a prototype is merely executable or genuinely deployment-ready and transferable across devices, sites, and seasons.

\vspace{-0.1in}
\section{Conclusion}
\vspace{-0.1in}

This study examined recent trends in deployment-oriented Edge AI and TinyML through the lens of resource-constrained computing, using precision farming as a representative environment in which cost, memory, energy, connectivity, and maintenance limitations directly influence system design. The four research questions were addressed through a common deployment evidence base and then synthesized by platform/toolchain (RQ1), optimization strategy (RQ2), intelligence placement (RQ3), and resource reporting (RQ4). The central finding across these dimensions is that deployment feasibility cannot be inferred from predictive accuracy or model size alone; it emerges from the interaction among target hardware, model architecture, runtime/compiler support, sensing and communication behavior, and the location of intelligence across endpoint, edge, and cloud tiers.

The analyzed systems show a heterogeneous hardware continuum, with microcontroller-class platforms such as ESP32, STM32, and ATmega serving as important endpoint targets, while single-board computers and embedded accelerators support workloads that exceed MCU-level memory or compute budgets. Across the optimization evidence, quantization and lightweight architecture design are the most mature and frequently used strategies. Structured pruning, multi-objective compression, hardware-aware neural architecture search, and explicit compiler-aware optimization remain less widely explored, despite their potential to optimize peak memory, measured latency, and energy rather than parameter count alone.

Resource profiling emerges as a primary methodological gap. Flash usage, peak SRAM, MAC or operation count, latency, power, and energy per inference are not reported consistently across studies. Because these quantities capture different aspects of deployability, future work should adopt standardized on-target profiling and clearly document the processor, numerical precision, runtime/toolchain, clock setting, and measurement protocol. This would enable reproducible comparison across devices and prevent nominally ``lightweight'' models from being evaluated solely through storage size or benchmark accuracy.

A second central insight is the inference--training asymmetry of contemporary Edge AI architectures. Inference is increasingly localized to sensing nodes, gateways, and mobile edge platforms, whereas training, retraining, model versioning, and lifecycle management remain predominantly centralized. This asymmetry provides autonomy and lower communication demand during inference, but it also motivates research on communication-efficient federated or split learning, lightweight adaptation, robust model-update mechanisms, and principled partitioning of intelligence across endpoint, edge, and cloud resources.

Taken together, the evidence suggests that the next generation of TinyML should be approached as a cross-layer systems discipline rather than simply as the execution of small models on small devices. Model compression, compiler/runtime optimization, sensing strategy, communication policy, scheduling, and power management need to be co-designed and evaluated against a common deployment budget. Precision farming provides a demanding testbed for this transition, but the resulting principles that is resource-aware profiling, hardware-software co-design, and communication-informed intelligence placement, are broadly applicable to resource-constrained Edge intelligence in other real-world domains.

\bibliographystyle{elsarticle-harv} 
\bibliography{example}






\end{document}